\documentclass[sigconf]{acmart}

\copyrightyear{2026}
\acmYear{2026}
\setcopyright{cc}
\setcctype{by}
\acmConference[CCS '26]{Proceedings of the 2026 ACM SIGSAC Conference on Computer and Communications Security}{November 15--19, 2026}{The Hague, Netherlands}
\acmBooktitle{Proceedings of the 2026 ACM SIGSAC Conference on Computer and Communications Security (CCS '26), November 15--19, 2026, The Hague, Netherlands}
\acmDOI{10.1145/3830454.3846557}
\acmISBN{979-8-4007-2871-6/2026/11}

\AtBeginDocument{%
  }
\usepackage{xspace}
\usepackage{tikz}
\usepackage{pifont}
\usepackage{textgreek}
\usepackage{multirow}
\usetikzlibrary{positioning, fit, arrows.meta, decorations.pathreplacing, calc}
\usepackage{cancel}
\usepackage{xcolor}

\newcommand{\drawcomments}{}

\newcommand{\newpar}[1]{\vspace{0.05cm}\noindent\textbf{#1.}\xspace}

\newcommand{\new}[1]{{\textcolor{blue}{~#1}}}

\ifdefined\drawcomments
  \newcommand{\zixu}[1]{\textcolor{blue}{[Zixu: #1]}}
  \newcommand{\david}[1]{\textcolor{teal}{[DL: #1]}}
\else
  \newcommand{\zixu}[1]{\ignorespaces}
  \newcommand{\david}[1]{\ignorespaces}
\fi

\newcommand{\toolname}{\textsc{PORTA}\xspace}

\begin{document}
\title{\toolname: Systematic Discovery of Vulnerabilities in Identity Brokers}

\author{Zixu Zhou}
\orcid{0009-0006-7536-9824}
\email{jasonz.zhou@utoronto.ca}
\affiliation{%
  \institution{University of Toronto}
  \city{Toronto}
  \state{Ontario}
  \country{Canada}
}

\author{Xiangyu Guo}
\email{xiangyu.guo@mail.utoronto.ca}
\orcid{0009-0000-9185-8296}
\affiliation{%
  \institution{University of Toronto}
  \city{Toronto}
  \state{Ontario}
  \country{Canada}
}

\author{Ilya Grishchenko}
\orcid{0000-0003-4744-7507}
\email{ilya.grishchenko@utoronto.ca}
\affiliation{%
  \institution{University of Toronto}
  \city{Toronto}
  \state{Ontario}
  \country{Canada}
}

\author{David Lie}
\orcid{0000-0002-2000-6827}
\email{david.lie@utoronto.ca}
\affiliation{%
  \institution{University of Toronto}
  \city{Toronto}
  \state{Ontario}
  \country{Canada}
}

\renewcommand{\shortauthors}{Zixu Zhou, Xiangyu Guo, Ilya Grishchenko, and David Lie}

\begin{abstract} 
Identity brokers such as Keycloak, Vault, and Dex translate authentication proofs from an upstream identity provider into sessions for downstream services. Due to their central role in authentication and trust mediation, a single vulnerability in a broker compromises every downstream service behind it. In fact, recent incidents have demonstrated that even widely adopted platforms are not immune to such issues, highlighting the critical need for systematic analysis.
To build such an analysis, this work first defines invariants for broker operation. These invariants are not completely covered by existing authentication specifications, primarily because brokers are unique in their duality, i.e., acting simultaneously as a relying party and an identity provider. The invariants capture the properties that identity brokers must uphold throughout their workflows, as all 60 previous CVEs we consider violate at least one invariant. 
Within these workflows, we further define high-level action categories. Equipped with this information, our automated tool \toolname leverages a large language model (LLM) to generate API invocation sequences corresponding to the high-level action categories, and oracle assertions aligned with the specified invariants.
\toolname identifies 33 previously unreported security vulnerabilities in seven popular open-source brokers. We study these novel vulnerabilities and trace them to eight recurring patterns.
We responsibly disclosed all discovered vulnerabilities: 29 have been confirmed, 14 patched, and 19 assigned CVE numbers, with 10 CVEs rated as critical.
\end{abstract}

\maketitle

\section{Introduction}\label{sec:intro}
During Single Sign-On (SSO), enterprises and cloud tenants do not authenticate users separately against each identity provider (IdP), each of which may rely on its own authentication protocol---such as Security Assertion Markup Language (SAML)~\cite{SAML-Core} for enterprise federation, OpenID Connect (OIDC)~\cite{OIDC-Core} for web and mobile SSO, and Open Authorization (OAuth)~\cite{RFC6749} for API authorization.
Instead, organizations typically operate an internal IdP that allows employees to authenticate once and subsequently reuse that authentication across different systems. 
This process involves relaying the initial authentication event across heterogeneous protocols.
Such relaying requires translating an authentication proof from one trust domain into a valid session within another.

The software that performs this translation is an
\emph{identity broker}~\cite{Keycloak,Dex,Vault,Authentik, Zitadel,Logto}.
Specifically, a broker receives an upstream authentication proof (e.g., a SAML assertion or an OIDC token), verifies its cryptographic integrity, extracts the identity and authorization claims it carries, maps those claims to a downstream principal, and issues a new credential for the downstream service.

Recent measurement work~\cite{InnocentiEtAl2025_BrokeredSSO} confirms that brokered SSO is pervasive on the internet
and that a single broker may sit in front of dozens
of relying parties (the applications and services that accept the broker's tokens) at once.

Naturally, a bug inside an identity broker affects every downstream service that trusts it. Moreover, the downstream services cannot detect or defend against the failure because by the time they see the broker's output, the upstream proof has already been replaced. The broker's internal credential-processing logic is therefore a single point of failure.

At the same time, broker operations violate a crucial assumption present in all authentication protocol specifications, namely, that each entity within an authentication protocol is assigned a single role. 
Consider a clause from the SAML specification~\cite{SAML-Core}: ``if \texttt{NotOnOrAfter} is present, the relying party \textsc{must} reject the assertion after that time.''
For a relying party (RP) that gets assertions from a single
trusted IdP, honoring the IdP's choice to omit the field is
correct.
For a broker, honoring the absent-field case enables an attacker to strip the field and disable the check.
Indeed, what distinguishes an identity broker from a conventional relying party or IdP is its \emph{dual role}: it simultaneously acts as a relying party to an upstream IdP and as an IdP to downstream clients.

The single role assumption is also assumed by existing analyses of authentication protocols by Fett et al.\ and
others~\cite{FettKuestersSchmitz2016_OAuthFormal,
FettKuestersSchmitz2017_OIDCFormal, FettHosseyniKuesters2019_FAPI}
which study security properties of the authentication protocols.
Other prior works either test single authentication protocol role (OAuch~\cite{PhilippaertsPreuveneersJoosen2022_OAuch} for OAuth authorization servers, COVScan~\cite{LuoEtAl2025_CrossAppOAuth} for app-platform OAuth) or study the external interaction between brokers and their environment~\cite{InnocentiEtAl2025_BrokeredSSO}. 

In this work, we study identity brokers as systems that combine an upstream relying party role with a downstream identity provider role.

First, we derive a set of five security invariants from a convergent analysis of 40 normative requirements across eight SSO specifications and 60 previous CVEs for ten broker platforms. Each invariant traces to at least one specification clause and is violated by CVEs in at least two independently developed brokers. We do not claim that these 5 invariants are a complete definition of correctness for broker operation, but rather that they are useful for detecting security vulnerabilities.


To detect invariant violations at scale, we design an automated approach with four stages: (1)~\emph{platform adaptation} uses an LLM to extract a platform profile from the broker's API documentation, mapping abstract operations to concrete HTTP endpoints and their parameters; (2)~\emph{sequence generation} uses an LLM to produce attack sequences and oracle assertions from the invariant definitions and the platform profile; (3)~\emph{execution} runs each sequence against a live broker instance in an isolated Docker container; and (4)~\emph{verdict} mechanically evaluates the oracle assertions against captured HTTP responses. 

We implement this approach as a tool we call \toolname, and evaluate it on seven broker platforms.

We first evaluated \toolname on a subset of the previous CVEs, selecting 11 applicable CVEs affecting the seven chosen broker platforms. Across three independent runs, \toolname found 10 of the 11 CVEs. 

We further conducted a vulnerability discovery campaign to identify previously unknown vulnerabilities. Using the same experimental setup, we performed three independent runs, each completing within 3 hours. \toolname produces 926 invariant violations before automatic deduplication. Deduplication within each run retains 252 violations, of which manual review confirms 226 and rejects 26. Source code analysis then maps the 226 confirmed findings to 33 distinct security flaws across the three runs. 15 of these flaws violate normative requirements in the authentication protocol specifications.
The remaining 18 are security-significant flaws where the invariant is violated but the relevant specification does not explicitly mandate a check, pointing to specification incompleteness in the dual-role broker context.

We studied newly identified vulnerabilities and observed similar failure patterns each recurring in two to six
independently developed broker codebases, written
in four programming languages by separate teams.
We attribute these recurrences to \emph{assumption mismatches}: vulnerable code preserves coding habits that are correct under the assumption that all parties are single-role, which is not the case for the dual-role identity brokers as discussed above.
In total, we identify eight distinct assumption-mismatch patterns and assign each of the 33 findings a dominant pattern.

From 33 findings, 15 (45\%) have root causes that do not match 
any CVE in the selected corpus, indicating that the invariant set generalizes beyond the CVEs from which it was derived.
We responsibly disclosed all violations to the affected maintainers; 29 have been confirmed, 14 patched, and 19 were assigned a CVE identifier. At least 10 of them were rated as critical.

In summary, the paper makes the following contributions:
\begin{itemize}
     \item We derive a set of five security invariants for identity broker dual-role operation from specifications and 60 previous CVEs.

    \item We design an automated approach that uses an LLM to generate sequences and oracle assertions from invariant definitions, and evaluates the assertions deterministically against captured platform behavior.

    \item We implement the approach as a tool \toolname and evaluate it on seven broker platforms, finding 10 of the 11 previously reported applicable security flaws and identifying 33 previously unreported security flaws.

    \item We categorize the discovered vulnerabilities into eight cross-platform failure patterns driven by a single role broker operation assumption. Responsible disclosure led to 29 acknowledgments, 14 patches, and 19 CVE assignments; at least 10 of them are critical. 

\end{itemize}

The source code of \toolname and the evaluation artifacts are available~\cite{porta}.
 
\section{Background}\label{sec:background}


\newpar{Identity Broker} An identity broker sits between an external Identity Provider (IdP) and
one or more downstream services.
Unlike a plain relying party that only \emph{consumes} an upstream
identity proof, a broker both consumes and \emph{re-issues}: it verifies
the upstream credential, extracts the identity and authorization claims
it carries, maps those claims to the downstream domain, and signs a
fresh credential for the downstream service.
This dual role, acting as a Service Provider (SP) toward the
upstream IdP and as an IdP toward downstream services, is what
makes brokers both powerful and fragile.
A failure in either half is directly exploitable: a verification mistake
lets an attacker forge upstream proofs, and an issuance mistake can hand
out downstream sessions with inflated privileges.
Platforms such as Keycloak, Authentik, and Dex are widely deployed in
this role, mediating enterprise SSO across dozens of downstream
apps~\cite{InnocentiEtAl2025_BrokeredSSO}.

\begin{figure*}[t]
\centering
\includegraphics[width=.9\textwidth]{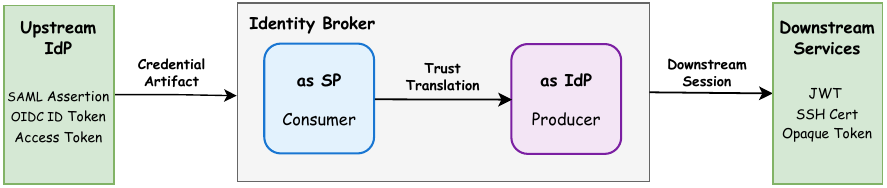}
\caption{An identity broker simultaneously acts as a Service Provider
  (SP/Consumer) toward the upstream IdP and as an Identity Provider
  (IdP/Producer) toward the downstream service. Security of the
  downstream session depends entirely on the correctness of the
  \emph{trust translation} step.}
\label{fig:broker-arch}
\end{figure*}

\newpar{Identity Broker Protocols} Brokering appears in two main protocol contexts.
In \emph{SAML brokering}, the broker receives a signed SAML~2.0 assertion from the upstream IdP, validates its signature and conditions, and issues its own downstream assertion or OIDC token~\cite{SAML-Core}. In \emph{OIDC and JWT federation}, the broker obtains an upstream ID token via the OAuth~2.0 Authorization Code
Flow~\cite{RFC6749} or validates a directly presented JSON Web Tokens (JWT)~\cite{RFC7519}, then issues a downstream session.
Token exchange~\cite{RFC8693} is a third pattern in which the broker translates one token type into another within a single protocol. The wire formats differ, but the security obligations are the same: the broker must not accept what it cannot verify, must not grant what was not authorized, and must not be manipulated through the sequencing of protocol steps. Figure~\ref{fig:broker-arch} illustrates the broker's position in the trust chain and its dual SP/IdP role. The shaded region between the upstream proof and the downstream session is the \emph{trust translation boundary}, which is the focus of this paper.

\section{Threat Model}
\label{sec:bg:threat}

\noindent\textbf{Attacker capabilities.}
We consider three attacker positions, each reflecting a realistic
scenario in federated identity deployments.
%
First, the attacker may \emph{operate an upstream IdP} that is
registered as a trusted identity source on the broker.
In multi-tenant broker deployments, any organization can register its
own IdP, so the attacker signs SAML assertions and JWTs with its own
keys and the broker is configured to trust those keys.
The attacker does not need to recover the cryptographic keys belonging to
other IdPs or forge their signatures; all attack credentials originate
from the attacker's own IdP.
Second, the attacker may \emph{intercept a legitimate upstream proof}
(e.g., via a compromised endpoint or a leaked URL) and attempt to
replay or modify it.
Third, the attacker may be a \emph{legitimate broker user} with
limited privileges who attempts to escalate access, for example by
exchanging a narrow-scope token for a broader one, or by accessing
resources of a different tenant.

\noindent\textbf{Out of scope.}
The scope of our approach is determined by several assumptions.
The broker is correctly configured with trust anchors for one or more
legitimate upstream IdPs; the attacker cannot read or modify the configuration, keys, or assertions of those legitimate IdPs.
The attacker cannot access the broker's database, file system, or internal network. The attacker cannot modify the broker's source code or configuration. 
All administrator configurations are trusted, and vulnerabilities due to administrator misconfigurations are out of scope. The attacker cannot access administrative APIs.
For instance, if an administrator explicitly enables a permissive setting (e.g., \texttt{trust-email=true} or \texttt{validateSignature=false}), the resulting behavior is by design,
not a vulnerability. Furthermore, we do not test denial-of-service conditions, client-side attacks (XSS, CSRF, clickjacking), or bugs in the upstream IdP itself.

\noindent\textbf{Deployment assumptions.}
We test each broker in its \emph{default configuration}, with at least one upstream IdP trust correctly provisioned by the administrator. Each test run starts from a clean state: the broker runs in an isolated container that is destroyed and recreated between protocol runs, ensuring no residual state from prior tests.

\noindent\textbf{Attacker goals.}
All goals are defined from the perspective of the downstream service
that trusts the broker's output.
The attacker succeeds if they can:
(a)~authenticate as a different principal (identity spoofing),
(b)~obtain a session that remains valid after the upstream proof
has expired or been revoked (replay),
(c)~gain permissions exceeding what the upstream IdP authorized
(privilege escalation), or
(d)~take over another user's account through identity collision
(account takeover).


\section{Trust Translation Invariants for Brokers}\label{sec:taxonomy}
To test whether identity brokers correctly translate trust between
upstream and downstream domains, we first need a precise definition
of what such correctness means for a broker.
Existing protocol specifications (e.g., SAML, OIDC, and OAuth) define
correctness for individual roles, namely what an IdP must assert
and what a relying party must verify, but not for the broker that
occupies both roles at once.
A direct consequence is that broker developers reading these
specifications can implement each role correctly in isolation and
still produce a translation pipeline that is insecure as a whole.
The goal of this section is to fill that gap with a set of correctness properties that (i)~are grounded in concrete normative clauses across all major SSO specifications, (ii)~are violated by real-world CVEs in multiple independently developed brokers, and (iii)~are testable through black-box API interaction.

We call these properties \emph{translation invariants}. 
The invariants serve three roles in the rest of the paper.
First, they make the broker's correctness requirements
\emph{explicit}, giving developers, auditors, and
standards bodies, a concrete checklist for broker testing.
Second, they organize the design of \toolname
(\S\ref{sec:design}): each invariant becomes a generation target
for attack sequences and a testing target for the verdict
pipeline.
Third, they provide a structural lens for interpreting our
findings (\S\ref{sec:eval:structural}): every cross-platform failure family we discover aligns with a specific invariant
boundary, which is direct evidence that the gap is structural
rather than incidental.

\subsection{Formal Model}
\label{sec:tax:model}

\begin{figure*}[t]
\centering
\includegraphics[width=\textwidth]{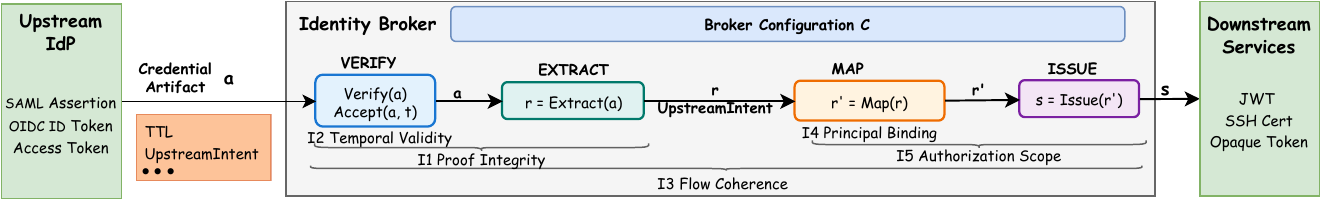}
\caption{The four-stage trust translation pipeline and where each
  invariant (\textbf{I1}--\textbf{I5}) applies.}
\label{fig:pipeline}
\end{figure*}

We model an identity broker as a trust translation function
\begin{equation}
  F \;:\; \mathcal{A}_{\mathrm{up}} \times \mathcal{C}
    \;\longrightarrow\;
    \mathcal{A}_{\mathrm{down}} \cup \{\bot\},
  \label{eq:F}
\end{equation}
where $\mathcal{A}_{\mathrm{up}}$ is the space of upstream credential
artifacts (e.g., SAML assertions, JWTs, and certificates), $\mathcal{A}_{\mathrm{down}}$ is the space of downstream sessions, $\mathcal{C}$ is the broker configuration (e.g., trust anchors, claim-mapping rules, and policy constraints), and $\bot$ denotes rejection.
$F$ decomposes into four sequential stages shown in Figure~\ref{fig:pipeline}:
\textsc{Verify} checks cryptographic validity and metadata against the trust anchors in~$\mathcal{C}$; \textsc{Extract} produces a structured representation of the verified artifact, including the principal identity, authentication context, and authorization claims; \textsc{Map} translates this representation to the downstream domain according to~$\mathcal{C}$'s mapping rules; and \textsc{Issue} serializes and signs the downstream credential.

Each boundary between these stages introduces a correctness requirement that does not arise in single-role IdP or RP implementations. We formalize these obligations as five security invariants. The function $F$ in Equation~(\ref{eq:F}) represents a single translation hop: one upstream IdP, one broker, and one downstream session. Real deployments often compose multiple hops, in which the output of one broker becomes the upstream input of the next. We restrict the formal treatment in this paper to the single-hop case and leave compositional analysis to future work. The five invariants we define apply to each hop independently and do not by themselves capture cross-hop properties such as transitive trust transfer.

\subsection{Derivation Methodology}
\label{sec:tax:method}

To derive a principled invariant set, we conducted a convergent analysis of two independent sources.

\noindent\textbf{Source~1: Protocol specs.}
We compiled 40 security-relevant normative requirements
(\textsc{must} and \textsc{shall} clauses) from eight
specifications: SAML~2.0 Core~\cite{SAML-Core} and
Profiles~\cite{SAML-Profiles},
OpenID Connect Core~1.0~\cite{OIDC-Core},
RFC~7519 (JWT)~\cite{RFC7519},
RFC~7515 (JWS)~\cite{RFC7515},
RFC~8693 (OAuth~2.0 Token Exchange)~\cite{RFC8693},
RFC~6749 (OAuth~2.0 Authorization Framework)~\cite{RFC6749},
and NIST SP 800-63C (Digital Identity Guidelines: Federation
and Assertions)~\cite{NIST80063C}.
These 40 clauses govern authentication proof
handling, session issuance, and authorization delegation, which correspond to the stages of $F$ operation.

\noindent\textbf{Source~2: Previous CVEs.}
We collected 60 broker logic CVEs for 10 identity broker platforms (Keycloak, Dex, Vault, Teleport, Ory Hydra/Fosite/Kratos, Conjur, Shibboleth, and OpenStack Keystone) via NVD, GitHub Security Advisories, and platform changelogs. The full list is in the artifact~\cite{porta}. We applied strict filtering consistent with our threat model (\S\ref{sec:bg:threat}): only authentication and authorization logic flaws were retained; XSS, CSRF, DoS, memory safety, and dependency bugs were excluded.
For each CVE, the violated requirement is determined by the \emph{fix commit itself}: the code change reveals which check was absent or incorrect, and this check maps directly to one of the 40 normative clauses above. 
For example, the fix for CVE-2018-14637 (Keycloak) adds \texttt{NotOnOrAfter} validation, mapping unambiguously to SAML~Core~\S2.5.1.2.

Grouping the 60 CVEs by their missing or incorrect security check produced five non-overlapping clusters that collectively account for all 60 CVEs. We refer to the properties violated by each cluster as \emph{invariants} which are formalized in~\S\ref{sec:tax:invariants}. We retain a cluster as an invariant only if it satisfies
three criteria: 1)~it is motivated by at least one citable clause in a protocol specification; 2)~it is violated by CVEs in at least two independently developed
platforms; and 3)~it is testable through black-box API interaction without source-code access.

\subsection{Invariant Definitions}
\label{sec:tax:invariants}
We now define each invariant with its formal specification, the pipeline stage it guards (Figure~\ref{fig:pipeline}), related protocol clauses, and a representative CVE.

We use two kinds of fonts. Broker operations, written in small caps, denote computations the broker actually performs on a credential. Additional functions we defined to concisely stipulate the invariants are in italic.
 
\noindent\textbf{I1: Proof Integrity} (\textsc{Verify} $\rightarrow$ \textsc{Extract}).
If the \textsc{Extract} stage finishes successfully when processing the credential artifact $a$, obtained from the successful \textsc{Verify} stage, then the representation $r$ derived from $a$ must match $a$ in content. Credential artifact $a$ and representation $r$ are records that include the following semantic context: the principal identity, the authentication context (e.g., authentication method and assurance level), and the authorization claims (e.g., group memberships).
We assume existence of an $SC$ function which outputs semantic context of $a$ and $r$.

\begin{equation}
  \textsc{Verify}(a) \wedge r = \textsc{Extract}(a) \implies \mathit{SC}(a) = \mathit{SC}(r)
  \label{eq:I1}
\end{equation}
This invariant is grounded in SAML~Core~\S5.4.2 (assertions must be integrity-protected), OIDC~Core~\S3.1.3.7 (signature and issuer validation), and RFC~7515~\S5.2 (JWS signature verification). I1 has two sub-patterns. \emph{I1-a (Signed $\neq$ Consumed):} the broker verifies one element but extracts identity from a different, unverified element. This is the classic XML Signature Wrapping attack~\cite{SomorovskyEtAl2012_BreakingSAML} and its JWT equivalents. \emph{I1-b (Wrong Verifier):} the broker applies the wrong verification policy, e.g., trusting an attacker-controlled JWKS endpoint or accepting a non-CA certificate as a trust anchor.

\emph{Representative CVE:} CVE-2024-8698 (Keycloak): the SAML signature validation logic verified one XML element, but the broker consumed identity claims from a different, unsigned element in the same response. An attacker can craft a response containing a legitimately signed element alongside an unsigned element carrying attacker-chosen identity claims; the broker verifies the former but grants a session reflecting the latter, enabling impersonation of any local account.

\noindent\textbf{I2: Temporal Validity}
(\textsc{Verify} stage).
During \textsc{Verify} stage, if the broker accepts $a$ at time $t + \Delta$ where $a$ was issued at time $t$, then $\Delta$ lies within $\mathit{TTL}(a)$. Let $k$ be the number of times the same artifact $a$ has been successfully accepted, including the current acceptance. If $a$ carries \texttt{OneTimeUse}, the broker may accept it only when $k=1$.

\begin{equation}
\begin{aligned}
  \textsc{Accept}(a, t + \Delta, k)
  \implies {}&
  0 \leq \Delta \leq \mathit{TTL}(a) \\
  & \land \bigl(\mathit{OneTimeUse}(a) \implies k=1\bigr)
\end{aligned}
  \label{eq:I2}
\end{equation}

This invariant is grounded in specifications SAML~Core~\S2.5.1.2 (\texttt{NotOnOrAfter}),
\S2.5.1.5 (\texttt{OneTimeUse}),
OIDC~Core~\S3.1.3.7 (\texttt{exp}, \texttt{nonce}), and also in RFC~7519~\S4.1.4 (\texttt{exp}).

\emph{Representative CVE:} CVE-2018-14637 (Keycloak): the SAML broker endpoint ignored assertion expiration conditions, allowing
replay of expired assertions to mint fresh downstream sessions.

\noindent\textbf{I3: Flow Coherence} \emph{(spans entire pipeline).}
If the broker issues a downstream session $s$ from representation $r'$, then $s$ must be derived by the complete execution of the pipeline.
There must exist an upstream artifact $a$ that the broker successfully verified, a representation $r$ that the broker extracted from $a$, and a downstream representation $r'$ that the broker mapped from $r$. The broker must not produce a session unless every preceding stage has succeeded on this same source artifact.

\begin{equation}
\begin{split}
  & s = \textsc{Issue}(r') \implies \\
  & \textsc{Verify}(a) \wedge
  r = \textsc{Extract}(a) \wedge 
  r' = \textsc{Map}(r)
\end{split}
\label{eq:I3}
\end{equation}

This rules out execution paths that skip a stage, accept state spliced from a different session, or bypass a configured authentication factor through an alternative entry point.

This invariant is not a property fully stated in existing specifications. Instead, it is a broker-specific generalization motivated by the following specification clauses: OIDC~Core~\S3.1.2.1 (\texttt{state}), \S3.3.2.6 (single-use authorization codes), SAML~Profiles~\S4.1.4.2~\cite{SAML-Profiles} (specifically, \texttt{InResponseTo} clause), and RFC~6749~\S4.1.3 (discussion on binding \texttt{client\_id}/\texttt{redirect\_uri}).

\emph{Representative CVE:} CVE-2021-3827 (Keycloak): the SAML ECP binding path bypassed other authentication flows including MFA, allowing an attacker to obtain a session without completing the required authentication steps.
 
\noindent\textbf{I4: Principal Binding} (\textsc{Map} stage).
If the principal identity component of two upstream representations $r_1, r_2$ are distinct under the broker's configuration (i.e., the configuration does not declare them as the same identity), the broker's \textsc{Map} stage must send them to distinct downstream identities $r_1', r_2'$.
We assume a function $\mathit{ID}$ which extracts the principal identity from representation.

\begin{equation} 
\begin{split}
  & \mathit{ID}(r_1)\neq  \mathit{ID}(r_2)
  \wedge
  r_1' = \textsc{Map}(r_1)
  \wedge r_2' = \textsc{Map}(r_2) \\
  & \implies
    \mathit{ID}(r_1')\neq  \mathit{ID}(r_2')
  \label{eq:I4}
\end{split}
\end{equation}
This invariant is grounded in OIDC~Core~\S2 (\texttt{sub} must uniquely identify the end-user), RFC~7519~\S4.1.2 (\texttt{sub}), and RFC~8693~\S2.1 (\texttt{subject\_token} identifies the acted-upon principal).

\emph{Representative CVE:} CVE-2020-16250 (Vault): the AWS IAM auth method accepted manipulated identity values, allowing one AWS principal to authenticate as a different one.

\noindent\textbf{I5: Authorization Scope} (\textsc{Map} $\rightarrow$ \textsc{Issue}). If the broker extracts representation $r$ from credential artifact $a$, maps $r$ to representation $r'$ (with $ID(r')$ principal), and issues session $s$ for $ID(r')$, then the permissions carried in $s$ are bounded by the upstream intent attached to $a$ and by what the broker's configuration grants to $ID(r')$.
The invariant stipulates that the granted permissions need to be aligned with both the original intent coming from the upstream IdP $\mathit{UpstreamIntent}$ and the broker configuration $\mathcal{C}$.
\begin{equation}
\begin{split}
  & r = \textsc{Extract}(a) \wedge r' = \textsc{Map}(r) \wedge s = \textsc{Issue}(r') \\
  & \implies \mathit{Perms}(s) \subseteq \mathit{UpstreamIntent} \cap \mathit{Policy}(\mathcal{C}, ID(r'))
\end{split}
\label{eq:I5}
\end{equation}
Here $\mathit{UpstreamIntent}$ denotes a set of permissions that the upstream intent wants to grant, they are encoded into the credential artifact $a$ that the broker receives as input. Further, we assume a function $Policy$, which takes as an input principal ID(r') and the broker configuration $\mathcal{C}$ and produces grounded instantiations of the generic permissions in $\mathcal{C}$.
Finally, we assume a function $Perms$ which extracts a set of permissions from the established session. As a concrete illustration, consider a JWT $a$ whose $\mathit{UpstreamIntent}$ wants to grant access to two APIs and specific actions within them: $\{(\text{finance-api}, \text{read:reports}),  (\text{comp-api}, \text{exec:task})\}$.
The broker in our example is configured such that 
$\mathit{Policy}(\mathcal{C}, ID(r'))$ returns $\{(\text{finance-api}, \text{read:reports}), (\text{hr-api}, \text{write:reports})\}$. For this example, the invariant I5 allows to grant $(\text{finance-api}, \text{read:reports})$ permission. By contrast, granting $(\text{comp-api}, \text{exec:task})$ is not permitted by the invariant as non-compliant with the policy, as well as the access to HR functionality $(\text{hr-api}, \text{write:reports})$, since it is not in the intent. 

This invariant is grounded in SAML~Core~\S2.5.1.4 (\texttt{Audience\-Restriction}), OIDC~Core~\S3.1.3.7 (\texttt{aud} must contain \texttt{client\_id}), RFC~7519~\S4.1.3 (\texttt{aud}), and RFC~8693~\S2.1 (\texttt{audience} and \texttt{scope} parameters).

\emph{Representative CVE:} CVE-2024-5798 (Vault): the JWT auth method did not validate the \texttt{aud} claim against the role-bound audience, allowing a token minted for one service to authenticate at another service.

\subsection{Coverage and Validation}
\label{sec:tax:coverage}

\begin{table}[t]
\centering
\small
\caption{Distribution of 60 CVEs across the five invariants.
  ``B \#'' counts independently developed broker codebases.
  ``S \#'' counts the normative requirements each invariant
  traces to across all eight specifications.}
\label{tab:invariant-dist}
\begin{tabular}{@{}llccl@{}}
\toprule
\textbf{Invariant} & \textbf{CVE} & \textbf{B \#} &
  \textbf{S \#} & \textbf{Example CVE} \\
\midrule
I1 Proof Integrity     & 19 & 7 & 10 & CVE-2024-8698  \\
I2 Temporal Validity   &  7 & 5 &  8 & CVE-2018-14637  \\
I3 Flow Coherence      & 10 & 4 &  9 & CVE-2021-3827 \\
I4 Principal Binding   &  6 & 2 &  6 & CVE-2020-16250  \\
I5 Authorization Scope & 18 & 4 &  7 & CVE-2024-5798  \\
\midrule
\textbf{Total}         & \textbf{60} & \textbf{10} & \textbf{40} & \\
\bottomrule
\end{tabular}
\end{table}
At least one of the five invariants is violated by each of the 60 CVEs in our dataset. Table~\ref{tab:invariant-dist} shows the invariant distribution over the 60 surveyed CVEs. Invariant I1 is the largest group (19~CVEs, 7~platforms), consistent with the long history of signature-related vulnerabilities in both SAML~\cite{SomorovskyEtAl2012_BreakingSAML} and JWT processing.
Invariant I5 is the second largest (18~CVEs), reflecting the widespread difficulty of enforcing scope constraints across token exchange and claim-mapping boundaries. Invariants I2 and I3 are smaller but span 4--5 platforms each. A potential concern is circularity: if the invariants are derived from CVEs, using them to find bugs may merely find those same CVEs. 

However, our evaluation (\S\ref{sec:eval:discovery}) shows that 45\% of the resulting findings match patterns with no prior CVE record, concentrated in the two invariants (I3 and I4) with the sparsest CVE coverage. We note that our invariants form only a partial specification of correctness for brokers, we do not claim they are complete.

\section{System Design}\label{sec:design}
\begin{figure*}[t]
\centering
\includegraphics[width=\textwidth]{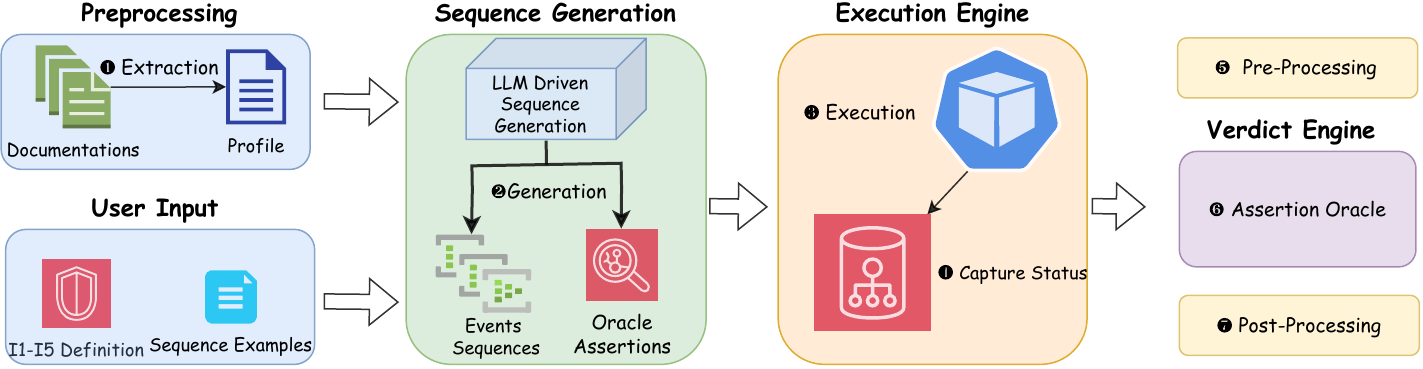}
\caption{System overview of \toolname pipeline.}
\label{fig:system-overview}
\end{figure*}

This section presents \toolname, a system that automatically tests identity brokers for violations of the invariants defined in~\S\ref{sec:taxonomy}. The core challenge is that each broker exposes a different API, yet the security properties to test (I1--I5) are platform-independent. \toolname addresses this with a four-phase pipeline (Figure~\ref{fig:system-overview}): (1)~\emph{platform adaptation} abstracts each broker's API into a common profile so that test logic is written once; 
(2)~\emph{sequence generation} uses an LLM to produce potential attack sequences and oracle assertions from the invariants;
(3)~\emph{execution} runs each sequence against a live broker instance; and (4)~\emph{verdict} evaluates captured behavior against the oracle to decide whether a violation occurred. The LLM participates only in Phases~1 and~2; Phases~3 and~4 are fully deterministic.

\subsection{Platform Adaptation}
\label{sec:des:profile}
Brokers under test can use different admin APIs, endpoint paths, and authentication flows. Writing test sequences directly against each API would require separate logic for every platform, and adding a new platform would mean writing new tests from scratch.
To address this challenge, \toolname{} introduces \emph{abstract operations}.
An abstract operation is a generic operation that must be instantiated for each platform. Although the instantiated operation's name and invocation mechanism may differ across platforms, the abstract operation produces the same semantic effect in each case, such as successfully creating a user account.  \toolname defines 33 abstract operations (e.g., \texttt{saml\_login}, \texttt{token\_exchange}, \texttt{create\_user}).\footnote{Note: it is pure coincidence that \toolname{} defines 33 abstract operations and happens to find 33 vulnerabilities.}
Nine of the 33 operations come from protocol specifications like RFC~6749 and the remaining 24 are administration operations shared by all seven popular broker platforms that we studied. The full list with specification references is in the artifact~\cite{porta}.

Having a set of abstract operations, our method automatically generates a \emph{platform profile} for each platform. 
The profile connects each abstract operation to a platform's HTTP method, URL path, request parameters, and response fields (Step~\ding{182} in Figure~\ref{fig:system-overview}). Specifically, for each platform, an LLM receives that platform's API documentation and produces the profile in a fixed structured format (see the artifact~\cite{porta} for the prompt skeleton). Because the LLM may produce incorrect endpoint paths or parameters, profile generation runs three times per platform and the three outputs are merged: each endpoint's HTTP method and path are decided by exact-string majority vote (at least two of three runs must agree), and body parameters and response fields are taken as a union. If all three runs disagree on a path, the operation is withheld and validated against the live broker before use. This case did not arise on any of the seven platforms we tested.

\subsection{Invariant-Guided Sequence Generation}
\label{sec:des:generation}
In the generation step, the LLM receives three inputs (Step~\ding{183} in Figure~\ref{fig:system-overview}): the platform profile, which lists the available abstract operations and how to call them on this platform; the target invariant definition (\S\ref{sec:tax:invariants}), which states the property the attack must violate; and one manually written example sequence per platform that we compose to illustrate the structural format the generator is expected to produce, including setup steps and execution steps. The example only provides the LLM with one generic sequence schema.

First, for each invariant I1--I5 and each platform, the LLM is asked to propose a pair of abstract operation sequences: a candidate attack sequence that probes this invariant for the given platform (e.g., replaying the same SAML assertion a second time to test I2), and compliant sequence that establishes the expected baseline (e.g., a correct SAML login that produces a valid downstream session).
Next, after obtaining abstract-operation sequences, the LLM is tasked to resolve them to platform-specific concrete HTTP requests using the profile (\S\ref{sec:des:profile}). This two-layer abstract/concrete architecture follows the model-based testing paradigm and has been used in security research to test multiple implementations of the same protocol~\cite{MaierEtAl2016_OAuthMBT, HussainEtAl2018_LTEInspector, ChenEtAl2023_SherlockOnSpecs}.

To avoid redundant execution, generated sequences are deduplicated before dispatch, e.g., sequences targeting the same invariant with structurally identical attack steps. Each campaign targets 10~sequences per invariant per protocol binding, yielding approximately 100 candidate sequences per platform before deduplication.

Alongside the sequences, the LLM produces an \emph{oracle assertion}. 
An oracle assertion is an instantiation of a generic invariant as a boolean expression over values captured from the executed sequence.
For example, the body of the I4 invariant $D(p_1, p_2) \Rightarrow \textsc{Map}(p_1) \neq \textsc{Map}(p_2)$ can be instantiated, for a sequence that logs in two distinct upstream principals through the same broker, as assertion \texttt{entity\_id\_first\_user != entity\_id\_second\_user. }

\subsection{Execution Engine}
\label{sec:des:execution}
After the platform-specific sequence and invariant resolution (\S\ref{sec:des:generation}), we execute LLM-generated sequences
against a live broker to produce evidence.
The Execution Engine runs each sequence against a live broker instance in an isolated Docker container
(Steps~\ding{184}--\ding{185} in Figure~\ref{fig:system-overview}). Before each run, the container is destroyed and recreated from a pinned image (Table~\ref{tab:platforms} lists all versions), ensuring a clean state with no residual sessions, tokens, or
configuration from prior tests.

During execution, the engine issues HTTP requests and chains values captured from one response into subsequent requests, for example feeding an authorization code returned by the authorize call into the following token exchange.

On each step that requires a credential, the engine produces it on demand. For SAML, the engine generates a signed assertion. For JWT, it mints a token whose standard claims defined in RFC7519\cite{RFC7519} (e.g., \texttt{iss}, \texttt{aud}, \texttt{exp}, and \texttt{sub}) take values the sequence specifies. Each credential is signed by a trusted key for the compliant sequence and by an attacker-controlled key for the attack sequence. Platform adapters handle protocol-specific flows such as OIDC authorization code exchange and multi-step IdP intent APIs without exposing these details to the sequence generator.

Transport errors and template-substitution failures abort the sequence and exclude it from all finding counts. A step in the compliant sequence that returns a non-200 status, however, still does not terminate the sequence. The co-generated oracle assertion observes the failed step and emits the corresponding finding for manual triage.

\subsection{Verdict Engine}
\label{sec:des:verdict}

\toolname's verdict engine evaluates oracle assertions. Each oracle assertion is a boolean expression, co-generated with the concrete sequences, that encodes what must be true if the invariant holds. 

Because the assertion is the per-sequence instantiation of the target invariant (\S\ref{sec:des:generation}), a false assertion corresponds to an invariant violation by the broker. The assertion oracle covers all five invariants and accounts for the majority of vulnerabilities discovered by \toolname.

\newpar{Pre-processing} \toolname has a pre-processing step before engaging the verdict engine. As each HTTP response arrives, \toolname applies lightweight syntactic checks that flag violations visible on a single response, without waiting for the sequence to complete (Step~\ding{186}). These include server errors (HTTP~5xx) on well-formed inputs, an issued JWT token whose \texttt{iss} claim does not match the broker's configured trusted issuer, and an OIDC \texttt{id\_token} missing the \texttt{nonce} that was authorized. This pre-processing reports only violations grounded directly in RFC-level requirements and contributes a small fraction of overall findings.

\newpar{Post-processing}
Apart from the verdict engine, \toolname has two post-processing sources of findings---fallback and mutation sweep.
For invariants I4 and I5, \toolname adds a fallback that does not rely on the LLM-generated assertions (Step~\ding{188}). It queries the broker's userinfo or introspection endpoint and compares the principal identity and the permissions in the upstream artifact against those carried by the session the broker actually granted. A mismatch in identity flags I4 and a mismatch in permissions flags I5.

\toolname also applies a manually curated library of 102 credential mutations (provided in the artifact~\cite{porta}, grouped by targeted invariant) to the concrete compliant sequence. We create this library, drawing from the 60 CVEs (\S\ref{sec:tax:method}) and documented SSO attack patterns such as XML Signature Wrapping~\cite{SomorovskyEtAl2012_BreakingSAML} and JWT tokens with the \texttt{alg=none} header that signals an unsigned token. 
Each mutation is applied individually so that acceptance of the mutated credential isolates which specific validation check the broker omits. Examples include removing a signature, changing an issuer, setting an expiration time in the past, replacing an audience, and stripping the SAML \texttt{<Conditions>} element. The library complements the testing with LLM-generated sequences. It contains mutations that aim to trigger violations of each of five invariants.

The mutation sweep reuses the compliant sequence produced in Phase~2 and changes one credential at a time. 
Each mutation targets a single element--users, providers, clients, organizations, policies, and authentication state---before performing a login, SAML callback, or token exchange. This lets \toolname{} test configuration changes, MFA paths, identity collisions, and token exchange across organizations.
\section{Evaluation}\label{sec:eval}
\subsection{Experimental Setup}
\label{sec:eval:setup}
We test seven open-source identity broker platforms across two protocol bindings (OIDC/JWT and SAML), yielding 13 distinct
platform/protocol configurations (Vault's SAML authentication is available only in the Enterprise edition; we test the open-source Community edition, which supports OIDC only). Table~\ref{tab:platforms} lists the platforms, versions, and GitHub stars as a proxy for deployment prevalence. Each platform runs in an isolated Docker container recreated from a pinned image between protocol runs.

We run three independent experiments, each using a separately generated set of platform profiles (\S\ref{sec:des:profile}), so that three-run variance reflects the full pipeline including LLM variation in profile generation.
Each experiment targets 10~sequences per invariant per protocol binding.

All LLM phases (profile generation, sequence generation, and oracle assertion generation) use a single configuration: \texttt{gpt-5\_2-2025-12-1}, default temperature, top-$p=1.0$. The three runs are fully automated from sequence generation to verdict assignment. 
Automatic finding deduplication is performed within each run. Identifying the source code relevant to the vulnerability-triggering sequence, consolidation into common root causes, and report preparation are performed manually across runs. 

\begin{table}[t]
\centering
\small
\caption{Target platforms tested by \toolname.}
\label{tab:platforms}
\begin{tabular}{@{}lllc@{}}
\toprule
\textbf{Platform} & \textbf{Version} & \textbf{Protocols} & \textbf{GitHub Stars} \\
\midrule
Keycloak~\cite{Keycloak}  & 26.5.6   & OIDC, SAML & 33.8k \\
Casdoor~\cite{Casdoor}   & 2.362.0  & OIDC, SAML & 13.3k \\
Dex~\cite{Dex}           & 2.45.1   & OIDC, SAML & 10.7k \\
Authentik~\cite{Authentik} & 2026.2.1 & OIDC, SAML & 20.9k \\
Vault~\cite{Vault} & 1.21.4 & OIDC & 35.4k \\
Zitadel~\cite{Zitadel}   & 4.13.0   & OIDC, SAML & 13.5k \\
Logto~\cite{Logto}       & 1.37.1   & OIDC, SAML & 11.9k \\
\bottomrule
\end{tabular}
\end{table}

\subsection{Recall on Known Vulnerabilities}
\label{sec:eval:redetection}

To measure recall, we evaluate \toolname{} on a subset of the 60-CVE corpus. In this corpus, 34 CVEs are for the applications in our test set (Keycloak, Dex, and Vault). Of those, 18 CVEs involve OIDC or SAML, the protocols that \toolname{} supports. However, seven of these require a protocol binding, concurrency, an external actor, or an operation that the executor currently does not support. This leaves 11 CVEs used for the recall experiment. The artifact~\cite{porta} gives details on the selection process. For each selected CVE, the recall experiment uses a pinned known-vulnerable, pre-patch version of the affected system, as documented in the artifact~\cite{porta}.

In the three independent experiments, \toolname generates a mean of $453.0 \pm 23.6$ sequences per run after removing structurally identical sequences before execution (\S\ref{sec:des:generation}).

We consider a CVE found only when a generated sequence triggers the behavior described in the CVE report. Table~\ref{tab:cve-redetection} reports 10 of 11, 7 of 11, and 8 of 11 CVEs found in the three experiments. The union detects 10 of 11 CVEs (90.9\%) for this subset and 10/18 (55.6\%) end-to-end when
all 18 OIDC and SAML CVEs are considered.
CVE-2024-5798 is detected only in one of the three rounds. The only CVE never detected is CVE-2018-14637. In this case, the generated tests were intended to use expired SAML assertions. However, the assertion builder set \texttt{NotOnOrAfter} 30 minutes after assertion creation, so the assertions were still valid when sent and the vulnerability was not triggered.
The API cost of the recall experiment is \$53.53.
\begin{table}[t]
\centering
\small
\caption{Detection of known CVEs in three independent experiments. Y and N indicate whether \toolname detected the CVE. The last column gives the number of experiments in which it was detected.}
\label{tab:cve-redetection}
\begin{tabular}{@{}lcccc@{}}
\toprule
\textbf{CVE} & \textbf{Exp.~1} & \textbf{Exp.~2} & \textbf{Exp.~3} & \textbf{Found} \\
\midrule
CVE-2018-14637 & N & N & N & 0/3 \\
CVE-2022-1245  & Y & N & Y & 2/3 \\
CVE-2023-0264  & Y & Y & Y & 3/3 \\
CVE-2023-2422  & Y & Y & Y & 3/3 \\
CVE-2023-2585  & Y & Y & Y & 3/3 \\
CVE-2023-6291  & Y & Y & N & 2/3 \\
CVE-2023-6544  & Y & Y & Y & 3/3 \\
CVE-2024-8698  & Y & N & Y & 2/3 \\
CVE-2025-1391  & Y & Y & Y & 3/3 \\
CVE-2024-5798  & Y & N & N & 1/3 \\
CVE-2025-3879  & Y & Y & Y & 3/3 \\
\midrule
\textbf{Detected} & \textbf{10/11} & \textbf{7/11} & \textbf{8/11} & \\
\bottomrule
\end{tabular}
\end{table}

\subsection{Vulnerability Discovery Results}
\label{sec:eval:discovery}

\begin{table}[t]
\centering
\small
\caption{Evaluation of \toolname on seven broker platforms across three independent experiments. \emph{Sequences}: \# of sequences generated in the experiment. \emph{Execution rate}: Percent of finished sequences. \emph{Findings:} \# of candidate vulnerabilities before deduplication. \emph{Deduplicated:} \# of candidate vulnerabilities after deduplication. \emph{True/False positives (TP/FP):} \# of vulnerabilities confirmed/rejected by manual triage. \emph{Precision:} $TP / (TP + FP)$.
}
\label{tab:precision}
\begin{tabular}{@{}lccc|c@{}}
\toprule
 & \textbf{Exp\,1} & \textbf{Exp\,2} & \textbf{Exp\,3} &
   $\boldsymbol{\mu \pm \sigma}$ \\
\midrule
Sequences            & 492 & 515 & 495 & $500.7 \pm 12.5$ \\
Execution rate       & 72.4\% & 69.5\% & 76.0\% & $72.6 \pm 3.3$\% \\
Findings     & 206 & 439 & 281 & $308.7 \pm 118.9$ \\
Deduplicated    &  54 &  96 & 102 & $84.0 \pm 26.2$ \\
True positives  &  44 &  89 &  93 & $75.3 \pm 27.2$ \\
False positives &  10 &   7 &   9 & $8.7 \pm 1.5$ \\
\textbf{Precision} & 81.5\% & 92.7\% & 91.2\% & $\mathbf{88.5 \pm 6.1}$\% \\
\bottomrule
\end{tabular}
\end{table}

In our three-run evaluation on the selected platforms (\S\ref{sec:eval:setup}), \toolname produces 10~sequences per invariant per protocol binding, with a mean of $500.7 \pm 12.5$ sequences per run after filtering structurally identical sequences  (\S\ref{sec:des:generation}).

The total API cost for all three runs is \$69.5.

Across three independent runs (Table~\ref{tab:precision}), \toolname detects 926 candidate vulnerabilities.
\toolname then uses an automatic deduplication deterministic pass that collapses findings sharing the following values: mutation rule + invariant (for mutation-sweep findings); invariant + verdict path + 
hash of the assertion text (for assertion violation findings). After deduplication, we obtain 252 candidate vulnerabilities ($84.0 \pm 26.2$ per run).

We attribute each candidate finding to a primary detection layer that fired first in the pipeline order. The verdict engine with assertion oracles (Step~\ding{187}) is the main mechanism with $70.0 \pm 21.2$ findings per run, about 83\% of all findings. The post-processing (Step~\ding{188}) has two components: the fallback contributes $7.0 \pm 4.6$ findings and the mutation sweep produces $4.0 \pm 1.7$ findings per run. The pre-processing  (Step~\ding{186}) contributes the remaining $3.0 \pm 1.0$ per run.

We performed manual analysis of 252 candidate vulnerabilities and confirm 226 of them, which corresponds to precision of $88.5 \pm 6.1\%$ with 26 false positives in total across the three runs. We further performed root-cause analysis and identified that 226 confirmed vulnerabilities are enabled by 33 distinct security flaws in the code discussed in \S\ref{sec:eval:structural}. We further identified that 26 false positives have four distinct root causes, the false-positive breakdown is in~\S\ref{sec:eval:precision}.

Table~\ref{tab:per-platform} summarizes the per-platform vulnerabilities breakdown.
We split all identified flaws into two categories: \emph{specification violations} and \emph{under-specified vulnerabilities}. This classification reflects specification strength, not exploitability: some under-specified flaws, such as identity collision via unverified email are easier to exploit in practice than some specification violations.

We also match each root cause vulnerability against the 60 CVEs: a finding is ``matched'' if any CVE's fix commit adds the same validation check that our finding shows is missing.

\newpar{Specification violations and under-specified vulnerabilities}
A flaw is a \emph{specification violation} if the observed behavior contradicts a normative \textsc{must} requirement in the specification.
For example, for temporal validity (I2), Authentik's JWT client-assertion endpoint accepts a token whose \texttt{exp} claim is ten minutes in the past (specification violation: the claim is present but ignored).
Of the 33 root-cause vulnerabilities, 15 are specification violations.
Examples include:
SAML assertion replay due to absent anti-replay caches
(I2, Keycloak, Casdoor, and Dex);
SAML AudienceRestriction not validated
(I5, Zitadel, Casdoor, and Authentik);
OIDC issuer verification disabled (I1, Authentik);
JWT client assertion expiration not enforced in the token
endpoint (I2, Authentik); and
OIDC nonce bypass via inverted check logic
(I2, Logto).

Union of the three runs covers all 15 specification violations, 12 of the 15 are found by at least two runs, and 5 of the 15 are found by all three runs. The three flaws found by only one run are two Authentik OIDC temporal-validity flaws (I2) found only in the first run (where the LLM generated the needed expiration-bypass sequences), and one Casdoor SAML proof-integrity flaw (I1) found only in the second run. 

\emph{Under-specified} flaws are insecure in practice but backed by optional or ambiguous normative language in the specification. 

For example, Zitadel's JWT handler skips expiration checking only when \texttt{exp} is absent (under-specified: RFC~7519 marks \texttt{exp} as optional). 

Of 33 identified root-cause vulnerabilities, 18 are under-specified.
Notable examples include: non-atomic configuration enforcement creating race windows (I3, Dex and Keycloak), cases where absent \texttt{<Conditions>} elements disable audience validation, a behavior the specification permits but does not recommend (I5, Keycloak and Dex), and identity collision through case-insensitive principal matching (I4, Vault).

Of the 18 under-specified flaws, all 18 are found by at least two runs and 9 of the 18 are found by all three runs, a higher rate than specification violations, reflecting that under-specified flaws are triggered by broader attack patterns that the LLM generates reliably across independent runs.

\begin{table}[t]
\centering
\small
\caption{Findings per platform. \emph{Spec.}: Specification violation (normative \textsc{must}).
  \emph{Under-spec.}: Security flaw is not covered by a specification.
  \emph{Matched}: Root-cause pattern matches an existing CVE on some platform in the dataset of \S\ref{sec:tax:method}.
  \emph{Non-matched}: Pattern does not match a CVE in the dataset.}
\label{tab:per-platform}
\begin{tabular}{@{}lccccc@{}}
\toprule
\textbf{Platform} & \textbf{Spec.} & \textbf{Under-spec.} & \textbf{Total}
  & \textbf{Matched} & \textbf{Non-matched} \\
\midrule
Keycloak  & 1 & 4 &  5 & 4 & 1 \\
Casdoor   & 6 & 3 &  9 & 5 & 4 \\
Dex       & 1 & 4 &  5 & 2 & 3 \\
Authentik & 4 & 1 &  5 & 2 & 3 \\
Vault     & 0 & 1 &  1 & 0 & 1 \\
Zitadel   & 2 & 1 &  3 & 2 & 1 \\
Logto     & 1 & 4 &  5 & 3 & 2 \\
\midrule
\textbf{Total} & \textbf{15} & \textbf{18} & \textbf{33}
  & \textbf{18} & \textbf{15} \\
\bottomrule
\end{tabular}
\end{table}



\noindent\textbf{Matched and non-matched patterns.}
During CVE matching, we compare each identified code defect against the 60-CVE corpus from \S\ref{sec:tax:method}. An example of a ``match'' is a pair of CVE-2018-14637's patch which fixes Keycloak by adding \texttt{NotOnOrAfter} enforcement and our finding that Casdoor also lacks \texttt{NotOnOrAfter} enforcement.
A finding is ``non-matched'' if no CVE fix in the dataset addresses the same problem. 
By this criterion, 15 of the 33 findings are non-matched and the other 18 are matched, although confirmed on platforms where they had not previously been reported (i.e., vulnerability ``transferred'' across the platforms). The non-matched count splits into 5 of the 15 specification violations and 10 of the 18 under-specified flaws. Non-matched findings concentrate in invariant I3 (six of seven) and I4 (two of four) violations, the two invariants with the sparsest CVE record
(Table~\ref{tab:invariant-dist}).

\noindent\textbf{Responsible disclosure.}
We began responsible disclosure to all affected platform maintainers on March 4, 2026. At the time of writing, 29 of the 33 disclosed flaws have been confirmed, 14 patched, and 19 assigned CVE identifiers. At least 10 of the assigned vulnerabilities have CVSS scores of 9.0 or higher, which is Critical severity.

\subsection{False Positive Analysis}
\label{sec:eval:precision}
The false-positive count per run (Table~\ref{tab:precision}) is stable ($8.7 \pm 1.5$, $\sigma/\mu = 17\%$).
We consolidate the 26 false positives into four distinct misclassification mechanisms.
Specifically, we group FPs by invariant violation which flagged it and the cause of misfire.

\newpar{Post-processing (10 FPs, I4, Step~\ding{188})} The I4 fallback (Step~\ding{188}, \S\ref{sec:des:verdict}) compares the principal in the upstream artifact against the principal the broker granted. For 10 cases on Casdoor, the broker returned the same user under a different display format (e.g., the sent username \texttt{i2\_logout\_user} versus the granted display name \texttt{I2 Logout User}), causing the string comparison to trigger even though no identity confusion occurred.

\newpar{Failed compliant sequence (8 FPs, multiple invariants, Step~\ding{187})} Sequence generation step (\S\ref{sec:des:generation}) produces a pair of attack and compliant sequences, both equipped with the oracle assertion checking the invariant. When the compliant sequence fails, i.e., returns a 4xx status (typically because the LLM used an incorrect endpoint or missing parameter), it may falsely trigger the oracle. For instance, a check whether the granted scope exceeds the requested one may falsely fire when the token request fails and no requested scope is captured.

\newpar{Pre-processing (4 FPs, I1, Step~\ding{186})} Four FPs are HTTP 500 server errors on malformed inputs (NUL bytes in headers, empty subject claims) that the per-response check (Step~\ding{186}) conservatively flagged as proof-integrity violations (I1). In each case the broker crashed during input parsing and never reached its authentication logic, so no invalid credential was accepted. 

\newpar{Test scaffold (4 FPs, Step~\ding{187})} The remaining four FPs arise from test infrastructure failures: Authentik returning an empty identity string and Vault rejecting a JWT due to a missing alias name. The oracle assertion (Step~\ding{187}) treats these infrastructure failures as invariant violations because the captured response values deviate from the expected.


\subsection{Ablation Study}
\label{sec:eval:ablation}

To measure how the design choices behind \toolname affect its results, we compare the full system against three configurations that turn off parts of the pipeline. These are configuration-level ablations rather than strict component-isolation experiments: each configuration changes a part of the pipeline, so the resulting deltas reflect what changing that part does to a whole system, not what a single part would do in isolation. Each configuration uses the same target platforms as the vulnerability discovery evaluation (\S\ref{sec:eval:discovery}), the same broker profile set, and three independent runs.

\begin{table*}
\centering
\small
\caption{Ablation results across three independent runs.
  }
\label{tab:ablation}
\begin{tabular}{@{}lllcccccc@{}}
\toprule
\textbf{Config} & \textbf{Generation} & \textbf{Verdict} &
  \textbf{Sequences/Run} & \textbf{Exec.\ Rate} &
  \textbf{Findings/Run} & \textbf{TP/Run} &
  \textbf{Precision} & \textbf{\# of findings} \\
\midrule
Blind          & generic prompt       & deterministic
  & $636.0 \pm 1.0$  & $56.5\pm 3.1\%$
  & $6.3 \pm 3.5$   & $0.0$           & $0.0\%$   & 0\,/\,33 \\
No-examples    & invariant only       & deterministic
  & $492.0 \pm 7.9$  & $63.3\pm 1.7\%$
  & $57.0 \pm 3.6$  & $43.0 \pm 6.0$  & $75.6\pm 11.5\%$  & 26\,/\,33 \\
LLM-judge      & invariant + examples & LLM
  & $475.3 \pm 7.0$  & $71.5\pm 2.2\%$
  & $245.3 \pm 22.7$ & $20.3 \pm 2.1$ & $8.4\pm 1.6\%$   & 12\,/\,33 \\
Full           & invariant + examples & deterministic
  & $500.7 \pm 12.5$ & $72.6\pm 3.3\%$
  & $84.0 \pm 26.2$ & $75.3 \pm 27.2$ & $88.5\pm 6.1\%$  & 33\,/\,33 \\
\bottomrule
\end{tabular}
\end{table*}

Table~\ref{tab:ablation} reports the aggregate results for the three ablation variants of the system.
By removing both the invariant definitions and the attack examples we obtain a Blind version of \toolname. It can use only the platform profile and a generic ``find security bugs'' prompt; the deterministic verdict pipeline remains intact. 
By keeping the invariant definitions in the generation prompt but removing the manually written examples, we obtain a No-example version.
LLM-judge version keeps the full invariant-guided generation but replaces the deterministic verdict pipeline with a single LLM call that decides whether each captured sequence shows a violation. The LLM-judge not only outputs whether the sequence is vulnerable, but also a confidence level for its verdict.
The three ablations isolate three independent dimensions: the invariant set determines what the system tests, the examples determine how precisely it tests, and the deterministic oracle determines whether the verdict can be trusted.

\newpar{Blind} 
The Blind version produces zero true positives across all three runs, and every one of its 19 findings is a false positive. 74\% of those are HTTP 500 server errors mislabeled as proof-integrity violations: with no target invariant, the LLM generates malformed inputs that crash request parsers rather than inputs that exercise authentication logic. This is not a sequence-count problem; Blind actually generates more sequences per run than the full system (636 vs.\ 501) but executes a smaller fraction of them successfully ($56.5\%$ vs.\ $72.6\%$), because the unguided sequences target the wrong endpoints with 
incorrect request formats.

\newpar{No-examples} 
No-examples version identifies 26 of 33 findings (a subset of findings in \S\ref{sec:eval:discovery}) on a similar sequence budget, but its execution rate drops from $72.6\%$ to $63.3\%$, per-run precision drops by 13 points ($88.5\%$ to $75.6\%$), and true positives per run drop by 43\% ($75.3$ to $43.0$). The seven findings (in \S\ref{sec:eval:discovery}) No-examples never  finds all require multi-step setup or non-obvious parameters, including cross-organization token exchange, the JWT client assertion temporal bypass, and identity collision. Invariant definitions alone are too abstract to steer the LLM toward these patterns.

\newpar{LLM-judge} Replacing the deterministic verdict pipeline with an LLM judge preserves the generation pipeline unchanged.
LLM-judge runs on a comparable number of sequences at a comparable execution rate, so any difference in outcome is attributable to the verdict layer rather than to the inputs.
Under these matched conditions, LLM-judge produces nearly three times as many findings per run as the full system
($245.3 \pm 22.7$ vs.\ $84.0 \pm 26.2$), but only
$20.3 \pm 2.1$ of those are true positives, for a precision of $8.4\%$ compared to the full system's $88.5\%$.
LLM-judge also misses 21 of the 33 included findings, including all of Casdoor's SAML audience-validation and assertion-replay findings and most of Authentik's I1 and I3 findings.
All 12 LLM-judge's findings are a subset of \toolname's findings in \S\ref{sec:eval:discovery}.

Manual triage of false positives reveals that $63\%$ of LLM findings are \emph{fabricated}, meaning the LLM reports concrete attack descriptions (e.g., ``intent-splicing bypass,'' ``application-tag enforcement missing'') that, on manual verification against the platform source, correspond to no actual security-relevant behavior. 
The deterministic assertion oracle cannot produce fabricated findings by construction: the assertion space is closed over typed-state boolean expressions written before execution, and every expression is evaluated mechanically against a concrete captured value.

Results in Table~\ref{tab:ablation} aggregate findings across all tested LLM confidence levels. Using a high-confidence setting for the LLM produces the best results, as precision reaches $38.2\%$, still a $2.3\times$ gap to the full system. In the artifact~\cite{porta}, we detail the results for different confidence configurations and whether adding I1--I5 invariant definitions to the judge prompt can improve the performance. All these configurations underperform compared to the one reported in Table~\ref{tab:ablation}.

\section{Security Findings and Insights}
\label{sec:eval:structural}


\subsection{Vulnerability Classification}
\label{sec:eval:classification}

This section studies whether real flaws reported in \S\ref{sec:eval:discovery} are isolated implementation mistakes, or they follow similar patterns.
Specifically, we analyze the 33 vulnerabilities along two axes: which security property is broken (failure families), 
and which coding assumption enabled the violation (assumption-mismatch patterns).
These axes are orthogonal: a family describes \emph{what} went wrong; a pattern describes \emph{why} the code was written that way.

\newpar{Cross-Platform Failure Families}
Table~\ref{tab:families} groups the 30 out of 33 findings by the broken security property and shows the seven most prevalent families,
each affecting two to six independently developed platforms.
Three of the 33 findings (Casdoor's cross-organization token
exchange, Casdoor's revoked-token acceptance, and Logto's nonce
predicate inversion) are single-instance, thus not aggregated into any family.

\begin{table*}[t]
\centering
\small
\caption{Cross-platform vulnerability families. \emph{Inv.}: Violated invariant.
  \emph{Platforms}: Number of independently developed platforms in the family.
  \emph{S/U}: Specification violations / Under-specified flaws within the family.
  \emph{Representative clause(s)}: Corresponding normative clause(s).
  \emph{Typical code form}: Recurring code construct that produces the flaw.
  }
\label{tab:families}
\begin{tabular}{@{}llccll@{}}
\toprule
\textbf{Family} & \textbf{Inv.} & \textbf{Platforms} &
  \textbf{S\,/\,U} & \textbf{Representative Clause(s)} &
  \textbf{Typical Code Form} \\
\midrule
Audience not validated
  & I5 & 6 & 4\,/\,4
  & RFC~7519 \S4.1.3; SAML \S2.5.1.4
  & \texttt{verify\_aud: False} or no audience check \\
Temporal validity
  & I2 & 4 & 3\,/\,2
  & SAML \S2.5.1.2; RFC~7519 \S4.1.4
  & \texttt{if field != nil \{ validate() \}} \\
Flow / non-atomic config
  & I3 & 4 & 0\,/\,5
  & ---
  & Config loaded at flow start, not re-checked \\
Assertion replay
  & I2 & 3 & 3\,/\,0
  & SAML \S2.5.1.5
  & No persistent assertion-ID cache \\
Issuer not validated
  & I1 & 3 & 2\,/\,1
  & OIDC \S3.1.3.7.4; SAML \S2.2.3
  & \texttt{verify\_iss: False} or cert from response \\
MFA bypass via federation
  & I3 & 2 & 1\,/\,1
  & NIST 800-63C~\S5.1~\cite{NIST80063C}
  & MFA check absent in federation code path \\
Identity collision
  & I4 & 3 & 0\,/\,4
  & ---
  & Case-insensitive or email-based linking \\
\bottomrule
\end{tabular}
\end{table*}

The family names describe the broken security property. 
Families violating invariants I1, I2, and I5 produce mostly specification violations (S) because those invariants have explicit \textsc{must} clauses in OIDC, SAML, and JWT specifications. Six of the seven I3 findings and all four I4 findings are under-specified flaws (U) that are not explicitly covered by specifications.

Typical code constructs related to the families (Table~\ref{tab:families}) are not instances of code reuse. The seven platforms are written in four different programming languages (Java, Go, Python, TypeScript) by independent development teams, and each family includes brokers built on top of unrelated third-party protocol libraries: Casdoor's SAML stack is built on \texttt{gosaml2}, Dex on \texttt{crewjam/saml}, Keycloak on its in-tree Java SAML core, Authentik on a custom Python SAML parser, Logto on \texttt{samlify}, Vault on the HashiCorp \texttt{cap/jwt}
plugin, and Zitadel on its own SAML SP implementation.


\begin{table*}
\centering
\small
\caption{\emph{Assumption-mismatch Patterns} by which single-role RP assumptions leak into dual-role broker code.
  \emph{\# of Findings}: The number of bugs that share the pattern.
  \emph{RP Assumption}: The defensive coding assumption that is correct for a single-role relying party.
  \emph{Broker Impact}: The consequence of the mismatch.}
\label{tab:patterns}
\begin{tabular}{@{}p{4.0cm}cp{5.5cm}p{5.5cm}@{}}
\toprule
\textbf{Assumption-mismatch Pattern} & \textbf{\# of Findings} &  \textbf{RP Assumption} & \textbf{Broker Impact} \\
\midrule
Silent skip on absent field
  & 9 & Honors IdP omission of optional field
  & Attacker strips field to disable check \\
\addlinespace
Explicit verification disabled
  & 8 & One trusted IdP makes the check redundant
  & Disabled check is the only namespace separator \\
\addlinespace
Stale configuration snapshot
  & 5 & Single-step request has no in-flight state
  & Multi-step flow spans configuration changes \\
\addlinespace
Cross-namespace identity collapse
  & 4 & Within one IdP, normalization is a convenience
  & Across IdPs, it merges distinct principals \\
\addlinespace
No persistent anti-replay state
  & 3 & Browser session is itself non-replayable
  & Broker accepts the same proof across sessions \\
\addlinespace
Branch coverage gap
  & 2 & Local-login path was hardened first
  & Federation path bypasses the local policy \\
\addlinespace
Trust bootstrap reflexivity
  & 1 & Cert is cached from out-of-band metadata
  & Cert is read from the proof being verified \\
\addlinespace
Non-state predicate
  & 1 & In trusted-IdP tests the two predicates agree
  & Attacker omits the field; predicate degenerates \\
\bottomrule
\end{tabular}
\end{table*}

\newpar{Assumption-Mismatch Patterns}
We observed that each newly identified vulnerability is enabled by an \emph{assumption mismatch}: 
a coding pattern that is correct for a single-role RP but becomes incorrect when the same code runs in a broker. 
Consider the scenario in which a single-role RP receives credentials from one pre-configured IdP that the operator has explicitly trusted. In this scenario, several 
coding practices are correct: honoring an IdP's choice to omit an optional field, caching the IdP certificate once at startup, validating only what the active code path needs, and treating the in-flight protocol state as a single transaction.
A broker code that applies these practices to upstream credentials treats attacker-supplied inputs as if they came from a pre-configured trusted source, which its threat model prohibits: the upstream IdP may be
attacker-controlled, the assertion may have been tampered with, the configuration may have changed mid-flow, or two upstream identities may collide in the broker's local namespace.

We label the 33 bugs by the form of the assumption mismatch. To do this, two authors independently labeled each finding with one dominant assumption pattern by reading the relevant source file (per-finding assignments and source links are in the artifact~\cite{porta}); disagreements in 3 cases were resolved by discussion (Cohen's $\kappa = 0.89$; all three disagreements were between silent skip and explicit disabling, the two most similar categories; details in the artifact~\cite{porta}).
The pattern set was not fixed in advance, and converged to the eight categories shown in Table~\ref{tab:patterns}. 
The two largest groups of findings by assumption-mismatch patterns---silent skip on absent field
(9 findings) and explicit verification disabled (8 findings)---together account for slightly over half of all findings.
Both arise directly from carrying RP-style ``honor what is given''
semantics into a context where what is given is
attacker-controlled.
The next largest are stale configuration snapshot (5 findings) and
cross-namespace identity collapse (4 findings). The I3 findings concern broker behavior over a complete flow, while the cited specifications state only the local checks within that flow. The I4 findings concern cross-IdP namespace isolation, which no single-role specification defines. These two groups therefore produce mostly under-specified flaws (\S\ref{sec:eval:catb}).


An assumption mismatch category is orthogonal to security family. 
A single family can mix multiple patterns (the audience family combines explicit disabling and silent skipping), and a single pattern can break multiple families (silent skipping breaks audience, temporal validity, and condition handling at once).
The full per-finding mapping appears in the artifact~\cite{porta}.

\subsection{Under-Specified Flaws}
\label{sec:eval:catb}

The specification-strength split across the 33 flaws reveals which invariant classes are least addressed by current specifications and practice. All four I4 flaws and six of seven I3 flaws are under-specified, meaning the violated requirement is either optional, undocumented, or debatable. The remaining I3 flaw violates a normative requirement clause. By contrast, I2 (temporal validity and replay) has seven corresponding specification violations out of ten, and I5 (audience and authorization scope) has five specification violations out of nine. This gradient reflects the degree of specification coverage: I1, I2, and I5 map to explicit \textsc{must} clauses, so violations are clear-cut. Contrastingly, the I3 invariant generalizes several normative clauses and is therefore not stated explicitly in the existing specifications. Consequently, most violations of this invariant are expected to be classified as under-specified. I4 corresponds to cross-IdP namespace isolation, which no specification addresses. These are the most structurally under-specified invariant classes in the identity broker ecosystem.

\subsection{Case Studies}
\label{sec:case-study}
We exemplify newly identified vulnerabilities via five case studies.
The first four are specification violations.
The fifth is an under-specified flaw presented as an end-to-end
attack scenario, included to show that under-specified does not
mean hard to exploit.
For each case, we summarize the broker's behavior, the attack
\toolname constructs, and the security impact.

\noindent\textbf{Case~A: JWT audience bypass (I5, Zitadel; assumption mismatch pattern: explicit verification disabled).}
Zitadel's Generic JWT IdP validates signature, issuer, and expiration of upstream JWTs, but never validates the audience claim against the configured downstream service. A JWT signed by the trusted IdP with \texttt{aud=service-B} authenticates successfully at a broker configured for
\texttt{service-A}.
\toolname constructs a control login with
\texttt{aud=service-A} (the service the broker is configured to protect) and an attack login with \texttt{aud=service-B} (a different service); the broker erroneously accepts both.

Impact: cross-service lateral movement, i.e., a token issued for one downstream service authenticates at every other service behind the same broker. 

This violates RFC~7519~\S4.1.3, which requires that each recipient processing a JWT verify that it appears in the \texttt{aud} claim.

\noindent\textbf{Case~B: Issuer verification disabled (I1,
Authentik; assumption mismatch pattern: explicit verification disabled).}
Authentik's OIDC source handler decodes upstream JWTs with
issuer verification turned off. A token from an attacker-controlled IdP with a mismatched \texttt{iss} claim is accepted as if it came from the configured trust anchor. \toolname constructs a legitimate login from the trusted IdP and an attack login signed by an attacker IdP, both with the same identity claims; the broker accepts both. 

Impact: complete trust anchor bypass, i.e., any issuer is treated as the configured trusted IdP. 

This violates OIDC Core~\S3.1.3.7.4, which requires that the \texttt{iss} claim in the token exactly match the provider's configured Issuer Identifier.

\noindent\textbf{Case~C: Nonce predicate inversion (I2, Logto;
assumption mismatch pattern: non-state predicate).}
Logto's OIDC SSO connector validates the nonce claim only when the
upstream token contains one, rather than whenever the session
sends one in the authorization request. An attacker, by omitting the nonce from the fabricated id\_token, bypasses the check entirely. \toolname generates an authentication that includes a nonce in the
request, then submits an id\_token with no \texttt{nonce} claim to the callback; the broker accepts the token. 

Impact: the nonce, which is intended to bind the id\_token to the authorization request and prevent replay, is rendered ineffective. 

This is a direct violation of OIDC Core~\S3.1.3.7.11.

\noindent\textbf{Case~D: MFA bypass via SAML federation (I3,
Casdoor; assumption mismatch pattern: branch coverage gap).}
Casdoor enforces multi-factor authentication for password and
OAuth login paths but does not invoke the MFA check on the SAML
federation path. A user enrolled in MFA can therefore complete login through a SAML upstream IdP without ever presenting the second factor. \toolname constructs an MFA-enrolled user, performs a control login through the password path (which correctly demands MFA), then performs the same login through the SAML path, which
returns a fully authenticated session in one step. 

Impact: MFA becomes voluntary, i.e., an attacker who recovers a user's password and a federated identity at a permissive upstream IdP can choose the federation path to bypass the local second-factor requirement. 

NIST~800-63C~\S5.1 requires that federation not be used to bypass authentication requirements.
 
\noindent\textbf{Case~E: Account takeover via unverified email
auto-linking (I4, Logto; under-specified).}
We include this case to show that an under-specified flaw can
correspond to a fully reproducible end-to-end attack.
Logto's SSO handler links an incoming SSO identity to an existing local account whenever the email matches, with no requirement that the upstream IdP have set \texttt{email\_verified}. We classify this as under-specified because no SSO specification normatively requires \texttt{email\_verified} for account linking. The attack proceeds through an upstream IdP that the administrator has already configured as a trusted SSO connector on the Logto tenant. The attacker registers \texttt{victim@corp.com} on that IdP (which does not verify email ownership) and logs in through it; Logto silently links the attacker-controlled SSO identity to the victim's existing local account, giving the attacker full session access on the next login. \toolname captures this by performing a password login as the victim, then submitting an SSO login from a mock IdP that asserts the victim's email without \texttt{email\_verified}, and observing that the resulting session resolves to the same local user identifier. 

Impact: account takeover for any local Logto account whose email is registrable on a permissive third-party IdP.

\subsection{Recommendations for Broker Specifications}

Based on the patterns reported in Table~\ref{tab:patterns}, we derive a set of high-level recommendations for broker specifications grouped into three categories: missing fields, configuration changes, and identity collisions.

\newpar{Missing fields} If the active broker policy requires a time, audience, nonce, or assertion condition check, an input that omits the required field must be rejected. Omission must not result in a skipped check. This recommendation follows from the I2 and I5 findings where validation is skipped when a field is absent.

\newpar{Configuration changes} If provider, trust, client, or MFA configuration changes before login completes, the broker must recheck the affected condition before issuing a session or reject the flow. 
This recommendation follows from the I3 findings where a login continues under an old configuration.

\newpar{Identity collisions} Identities from different upstream IdPs must remain distinct unless an explicit linking rule declares them to be the same account. If email is used for linking, the rule must require a verified email claim. This requirement follows from the I4 identity collision findings and the Logto case study (\S\ref{sec:case-study}).
\section{Discussion and Limitations}\label{sec:limitations}
\newpar{Hypothesis coverage of the sequence generator}

The sequence generator samples attack strategies instead of enumerating every way to violate an invariant.

For example, in one sequence, where the mapper normalizes email casing inconsistently, the oracle may be triggered for I4 because two accounts are mapped to the same identity. 

However, for the same broker, \toolname{} may not identify a scenario that triggers the same I4 invariant through a user identifier collision.

\newpar{Recall experiment}
As the experiment demonstrates the detection rate on a subset of known CVEs, it can serve only as a lower-bound estimate of the false-negative rate of \toolname{}. In addition, the current instantiation of \toolname{} has limitations due to implementation choices.

Specifically, CVE-2018-14637 requires an expired assertion, which is not currently supported because \toolname{}'s builder always generates an assertion with a future expiration time.

\newpar{Invariant scope}
The five invariants were derived from the representative set of 60 CVEs. Another corpus may identify additional broker requirements or organize the same cases differently. \toolname tests the five invariants defined in this paper, but these invariants may not cover every possible broker security property.

\newpar{Default-configuration scope}
All findings are relative to pinned, minimally configured deployments. Enterprise installations routinely introduce custom attribute mappers, claim transformation scripts, external user stores, and group membership plugins. A platform that satisfies all five invariants under default configuration may fail them after customization. Our results characterize the translation logic each broker platform ships with, not the security of any specific production deployment.

\section{Related Work}\label{sec:related}
\newpar{Protocol-level formal analysis}
A substantial body of work establishes formal security guarantees for
SSO protocol roles.
Fett et al
produced comprehensive models of
OAuth~2.0~\cite{FettKuestersSchmitz2016_OAuthFormal} and
OpenID Connect~\cite{FettKuestersSchmitz2017_OIDCFormal} using the
Web Infrastructure Model, later extended to the Financial-grade
API~\cite{FettHosseyniKuesters2019_FAPI} and
FAPI~2.0~\cite{HosseyniKuestersWuertele2024_FAPI2}.
On the SAML side, Gro\ss{} analyzed the Browser/Artifact
profile~\cite{Gross2003_SAMLBrowserArtifact}, and Armando et al.\
formally analyzed and broke the SAML-based SSO flow for Google
Apps~\cite{ArmandoEtAl2008_SAMLFormal}.
These results define security properties for individual protocol roles
(IdP, RP, authorization server, client) under adversary models that
assume correct role behavior.
They do not model what happens when a \emph{broker} simultaneously
occupies multiple roles, nor do they define the correctness obligations
that arise at the verify--extract--map--issue broker translation boundaries.
Our work targets exactly this gap: we ask what invariants
a dual-role broker must preserve as it translates upstream proofs into downstream
sessions, across protocol boundaries.

\newpar{Empirical SSO security and implementation flaws}
Traffic-guided studies by Wang et al.~\cite{WangChenWang2012_WebSSO}
and Sun and Beznosov~\cite{SunBeznosov2012_OAuthImplDetails}
demonstrated that commercially deployed OAuth and OpenID
implementations are rife with logic errors.
Automated RP-side testing tools such as
SSOScan~\cite{ZhouEvans2014_SSOScan} and
SAAT~\cite{GhasemisharifKanichPolakis2022_SAAT} later confirmed
that integration failures persist years after initial disclosure.
Ghasemisharif et al.~\cite{GhasemisharifEtAl2018_SingleSignOff}
took a complementary empirical approach, measuring SSO account
hijacking and session-management failures across deployed websites.
Somorovsky et al.'s XML Signature Wrapping
work~\cite{SomorovskyEtAl2012_BreakingSAML} is a direct technical
ancestor of our I1 (Proof Integrity) invariant: it proved that the
``verified object $\neq$ consumed object'' failure mode is systemic
across SAML implementations.
Our work extends this line in two directions: we generalize beyond a
single invariant and single protocol to five generic invariants,
and we shift the analysis unit from individual RP or IdP implementations
to the broker's translation pipeline.

\newpar{IdP testing, compliance, and malicious-IdP attacks}
Mainka, Mladenov, and Schwenk~\cite{MainkaMladenovSchwenk2016_MaliciousIdP}
introduced the malicious-IdP attacker model and used it to study
SSO clients across OpenID and OpenID Connect deployments.
Their subsequent SoK~\cite{MainkaMladenovSchwenkWich2017_SoK_OIDC}
systematized OIDC security properties and their evaluation.
OAuch~\cite{PhilippaertsPreuveneersJoosen2022_OAuch} tests OAuth
authorization servers against standards and best-practice requirements,
demonstrating the value of executable compliance oracles.
Philippaerts et al.\ later extended this approach to evaluate
open-source identity provider middleware~\cite{PhilippaertsEtAl2024_OAuthMiddleware}.
Bisegna et al.~\cite{BisegnaEtAl2024_MicroIdGym} introduced
Micro-Id-Gym for testing identity management implementations, and
Yang et al.~\cite{YangEtAl2026_TokenTimeBomb} fuzzed JWT library
implementations for temporal and signature validation flaws.
\toolname shares the compliance-oracle philosophy with OAuch, but
targets the cross-protocol, cross-role translation pipeline rather
than single-role protocol compliance.
OAuch's checklist is built around an authorization server's
obligations to a single relying party, so it has no notion of an
upstream IdP feeding the system under test and cannot test
multi-step flow atomicity or cross-IdP namespace isolation.
The 11 of our 33 findings that violate I3 and
I4 invariants are precisely the classes OAuch's
compliance model has no representation for.

\newpar{Brokered and intermediary SSO}
The closest prior work targets the broker layer directly.
Innocenti et al.~\cite{InnocentiEtAl2025_BrokeredSSO} study
brokered SSO on the public Web, quantifying broker deployment prevalence
and identifying vulnerabilities in the \emph{external} redirect-chain
composition between brokers, upstream IdPs, and downstream sites.
Luo et al.~\cite{LuoEtAl2025_CrossAppOAuth} introduce cross-app
token attacks and the COVScan tester for OAuth~2.0 
platforms; the intermediary layer is superficially similar, but the
target is app-to-platform account linking rather than the intra-broker
translation chain.
Jannett et al.'s DISTINCT attack~\cite{JannettEtAl2022_DISTINCT}
exploits in-browser communication between dual SSO windows, targeting
the browser-side composition rather than the server-side broker logic.
Our work is complementary: where Innocenti et al.\ study how brokers
are \emph{composed} with their environment, we study whether the
broker's \emph{internal} translation logic preserves the five
invariants we define.
To our knowledge, no prior work provides a systematic,
implementation-grounded specification of the correctness obligations
that arise inside the broker's translation pipeline across multiple
platforms and protocol bindings.

\section{Conclusion and Future Work}\label{sec:conclusion}

Identity brokers implement a trust translation function between IdPs and downstream services.

In this work, we formalized five translation invariants (I1--I5) from 40 normative protocol requirements and 60 previous CVEs, and built \toolname to test whether deployed brokers satisfy them. 

We consolidated \toolname's findings into 33 previously unreported broker security flaws. At the time of revision, 29 have been confirmed, 14 patched, and 19 assigned CVE identifiers. At least 10 of the assigned vulnerabilities have critical severity.

For future work, we plan to engage with standards bodies to lift the I3 and I4 invariants into specifications, as most of their violations identified by our tool are not currently covered. Further,  we plan to extend the translation model from single brokers to chains of brokers so that invariant violations can be studied as they propagate across multiple translation hops.
\section*{Acknowledgements}

Funding for this work was provided in part by NSERC Discovery Grant RGPIN-2026-07548 and NSERC-CSE Research Communities Grant ALLRP 588144-23. Credits for LLM access were generously provided by OpenAI. David Lie is supported by Tier 1 Canada Research Chair CRC-2019-00242 and the Schwartz Reisman Institute for Technology and Society. Researchers funded through the NSERC-CSE Research Communities Grants do not represent the Communications Security Establishment Canada or the Government of Canada. Any research, opinions or positions they produce as part of this initiative do not represent the official views of the Government of Canada.


\bibliographystyle{ACM-Reference-Format}
\balance
\bibliography{references}

@inproceedings{WangChenWang2012_WebSSO,
author = {Wang, Rui and Chen, Shuo and Wang, XiaoFeng},
title = {Signing Me onto Your Accounts through Facebook and Google: A Traffic-Guided Security Study of Commercially Deployed Single-Sign-On Web Services},
year = {2012},
isbn = {9780769546810},
publisher = {IEEE Computer Society},
address = {USA},
url = {https://doi.org/10.1109/SP.2012.30},
doi = {10.1109/SP.2012.30},
booktitle = {Proceedings of the 2012 IEEE Symposium on Security and Privacy},
pages = {365–379},
numpages = {15},
series = {SP '12}
}

@inproceedings{SunBeznosov2012_OAuthImplDetails,
author = {Sun, San-Tsai and Beznosov, Konstantin},
title = {The devil is in the (implementation) details: an empirical analysis of OAuth SSO systems},
year = {2012},
isbn = {9781450316514},
publisher = {Association for Computing Machinery},
address = {New York, NY, USA},
url = {https://doi.org/10.1145/2382196.2382238},
doi = {10.1145/2382196.2382238},
booktitle = {Proceedings of the 2012 ACM Conference on Computer and Communications Security},
pages = {378–390},
numpages = {13},
location = {Raleigh, North Carolina, USA},
series = {CCS '12}
}

@inproceedings{ZhouEvans2014_SSOScan,
author = {Zhou, Yuchen and Evans, David},
title = {SSOScan: automated testing of web applications for single sign-on vulnerabilities},
year = {2014},
isbn = {9781931971157},
publisher = {USENIX Association},
address = {USA},
booktitle = {Proceedings of the 23rd USENIX Conference on Security Symposium},
pages = {495–510},
numpages = {16},
location = {San Diego, CA},
series = {SEC'14}
}

@INPROCEEDINGS{GhasemisharifKanichPolakis2022_SAAT,
  author={Ghasemisharif, Mohammad and Kanich, Chris and Polakis, Jason},
  booktitle={2022 IEEE Symposium on Security and Privacy (SP)}, 
  title={Towards Automated Auditing for Account and Session Management Flaws in Single Sign-On Deployments}, 
  year={2022},
  volume={},
  number={},
  pages={1774-1790},
  doi={10.1109/SP46214.2022.9833753}}

@inproceedings{GhasemisharifEtAl2018_SingleSignOff,
author = {Ghasemisharif, Mohammad and Ramesh, Amruta and Checkoway, Stephen and Kanich, Chris and Polakis, Jason},
title = {O single sign-off, where art thou? an empirical analysis of single sign-on account hijacking and session management on the web},
year = {2018},
isbn = {9781931971461},
publisher = {USENIX Association},
address = {USA},
booktitle = {Proceedings of the 27th USENIX Conference on Security Symposium},
pages = {1475–1492},
numpages = {18},
location = {Baltimore, MD, USA},
series = {SEC'18}
}

@inproceedings{FettKuestersSchmitz2016_OAuthFormal,
author = {Fett, Daniel and K\"{u}sters, Ralf and Schmitz, Guido},
title = {A Comprehensive Formal Security Analysis of OAuth 2.0},
year = {2016},
isbn = {9781450341394},
publisher = {Association for Computing Machinery},
address = {New York, NY, USA},
url = {https://doi.org/10.1145/2976749.2978385},
doi = {10.1145/2976749.2978385},
booktitle = {Proceedings of the 2016 ACM SIGSAC Conference on Computer and Communications Security},
pages = {1204–1215},
numpages = {12},
location = {Vienna, Austria},
series = {CCS '16}
}

@INPROCEEDINGS{FettKuestersSchmitz2017_OIDCFormal,
  author={Fett, Daniel and Küsters, Ralf and Schmitz, Guido},
  booktitle={2017 IEEE 30th Computer Security Foundations Symposium (CSF)}, 
  title={The Web SSO Standard OpenID Connect: In-depth Formal Security Analysis and Security Guidelines}, 
  year={2017},
  volume={},
  number={},
  pages={189-202},
  doi={10.1109/CSF.2017.20}}

@INPROCEEDINGS{FettHosseyniKuesters2019_FAPI,
  author={Fett, Daniel and Hosseyni, Pedram and Küsters, Ralf},
  booktitle={2019 IEEE Symposium on Security and Privacy (SP)}, 
  title={An Extensive Formal Security Analysis of the OpenID Financial-Grade API}, 
  year={2019},
  volume={},
  number={},
  pages={453-471},
  doi={10.1109/SP.2019.00067}}

@article{HosseyniKuestersWuertele2024_FAPI2,
author = {Hosseyni, Pedram and K\"{u}sters, Ralf and W\"{u}rtele, Tim},
title = {Formal Security Analysis of the OpenID FAPI 2.0 Family of Protocols: Accompanying a Standardization Process},
year = {2024},
issue_date = {February 2025},
publisher = {Association for Computing Machinery},
address = {New York, NY, USA},
volume = {28},
number = {1},
issn = {2471-2566},
url = {https://doi.org/10.1145/3699716},
doi = {10.1145/3699716},
journal = {ACM Trans. Priv. Secur.},
month = nov,
articleno = {4},
numpages = {36}
}

@inproceedings{SomorovskyEtAl2012_BreakingSAML,
author = {Somorovsky, Juraj and Mayer, Andreas and Schwenk, J\"{o}rg and Kampmann, Marco and Jensen, Meiko},
title = {On breaking SAML: be whoever you want to be},
year = {2012},
publisher = {USENIX Association},
address = {USA},
booktitle = {Proceedings of the 21st USENIX Conference on Security Symposium},
pages = {21},
numpages = {1},
location = {Bellevue, WA},
series = {Security'12}
}

@inproceedings{Gross2003_SAMLBrowserArtifact,
author = {Groβ, Thomas},
title = {Security Analysis of the SAML Single Sign-on Browser/Artifact Profile},
year = {2003},
isbn = {0769520413},
publisher = {IEEE Computer Society},
address = {USA},
booktitle = {Proceedings of the 19th Annual Computer Security Applications Conference},
pages = {298},
series = {ACSAC '03}
}

@inproceedings{ArmandoEtAl2008_SAMLFormal,
author = {Armando, Alessandro and Carbone, Roberto and Compagna, Luca and Cuellar, Jorge and Tobarra, Llanos},
title = {Formal analysis of SAML 2.0 web browser single sign-on: breaking the SAML-based single sign-on for google apps},
year = {2008},
isbn = {9781605582887},
publisher = {Association for Computing Machinery},
address = {New York, NY, USA},
url = {https://doi.org/10.1145/1456396.1456397},
doi = {10.1145/1456396.1456397},
booktitle = {Proceedings of the 6th ACM Workshop on Formal Methods in Security Engineering},
pages = {1–10},
numpages = {10},
location = {Alexandria, Virginia, USA},
series = {FMSE '08}
}

@INPROCEEDINGS{InnocentiEtAl2025_BrokeredSSO,
  author={Innocenti, Tommaso and Jannett, Louis and Mainka, Christian and Mladenov, Vladislav and Kirda, Engin},
  booktitle={2025 IEEE Symposium on Security and Privacy (SP)}, 
  title={“Only as Strong as the Weakest Link”: On the Security of Brokered Single Sign-On on the Web}, 
  year={2025},
  volume={},
  number={},
  pages={1009-1027},
  doi={10.1109/SP61157.2025.00024}}

@inproceedings{LuoEtAl2025_CrossAppOAuth,
author = {Luo, Kaixuan and Wang, Xianbo and Fung, Pui Ho Adonis and Lau, Wing Cheong and Lecomte, Julien},
title = {Universal cross-app attacks: exploiting and Securing OAuth 2.0 in integration platforms},
year = {2025},
isbn = {978-1-939133-52-6},
publisher = {USENIX Association},
address = {USA},
booktitle = {Proceedings of the 34th USENIX Conference on Security Symposium},
articleno = {166},
numpages = {18},
location = {Seattle, WA, USA},
series = {SEC '25}
}

@inproceedings{PhilippaertsPreuveneersJoosen2022_OAuch,
author = {Philippaerts, Pieter and Preuveneers, Davy and Joosen, Wouter},
title = {OAuch: Exploring Security Compliance in the OAuth 2.0 Ecosystem},
year = {2022},
isbn = {9781450397049},
publisher = {Association for Computing Machinery},
address = {New York, NY, USA},
url = {https://doi.org/10.1145/3545948.3545955},
doi = {10.1145/3545948.3545955},
booktitle = {Proceedings of the 25th International Symposium on Research in Attacks, Intrusions and Defenses},
pages = {460–481},
numpages = {22},
location = {Limassol, Cyprus},
series = {RAID '22}
}

@INPROCEEDINGS{MainkaMladenovSchwenk2016_MaliciousIdP,
  author={Mainka, Christian and Mladenov, Vladislav and Schwenk, Jörg},
  booktitle={2016 IEEE European Symposium on Security and Privacy (EuroS\&P)}, 
  title={Do Not Trust Me: Using Malicious IdPs for Analyzing and Attacking Single Sign-on}, 
  year={2016},
  volume={},
  number={},
  pages={321-336},
  doi={10.1109/EuroSP.2016.33}}

@INPROCEEDINGS{MainkaMladenovSchwenkWich2017_SoK_OIDC,
  author={Mainka, Christian and Mladenov, Vladislav and Schwenk, Jörg and Wich, Tobias},
  booktitle={2017 IEEE European Symposium on Security and Privacy (EuroS\&P)}, 
  title={SoK: Single Sign-On Security — An Evaluation of OpenID Connect}, 
  year={2017},
  volume={},
  number={},
  pages={251-266},
  doi={10.1109/EuroSP.2017.32}}

@inproceedings{JannettEtAl2022_DISTINCT,
author = {Jannett, Louis and Mladenov, Vladislav and Mainka, Christian and Schwenk, J\"{o}rg},
title = {DISTINCT: Identity Theft using In-Browser Communications in Dual-Window Single Sign-On},
year = {2022},
isbn = {9781450394505},
publisher = {Association for Computing Machinery},
address = {New York, NY, USA},
url = {https://doi.org/10.1145/3548606.3560692},
doi = {10.1145/3548606.3560692},
booktitle = {Proceedings of the 2022 ACM SIGSAC Conference on Computer and Communications Security},
pages = {1553–1567},
numpages = {15},
location = {Los Angeles, CA, USA},
series = {CCS '22}
}

@INPROCEEDINGS{PhilippaertsEtAl2024_OAuthMiddleware,
  author={Philippaerts, Pieter and Vanhoof, Jan and Van Cutsem, Tom and Joosen, Wouter},
  booktitle={GLOBECOM 2024 - 2024 IEEE Global Communications Conference}, 
  title={Is Your OAuth Middleware Vulnerable? Evaluating Open-Source Identity Providers’ Security}, 
  year={2024},
  volume={},
  number={},
  pages={3607-3612},
  doi={10.1109/GLOBECOM52923.2024.10901500}}

@article{BisegnaEtAl2024_MicroIdGym,
author = {Bisegna, Andrea and Bitussi, Matteo and Carbone, Roberto and Ranise, Silvio},
title = {Enhancing Security Testing for Identity Management Implementations: Introducing Micro-Id-Gym Language and Micro-Id-Gym Testing Tool},
year = {2024},
issue_date = {Nov.-Dec. 2024},
publisher = {IEEE Educational Activities Department},
address = {USA},
volume = {22},
number = {6},
issn = {1540-7993},
url = {https://doi.org/10.1109/MSEC.2024.3450277},
doi = {10.1109/MSEC.2024.3450277},
journal = {IEEE Security and Privacy},
month = nov,
pages = {50–61},
numpages = {12}
}

@inproceedings{YangEtAl2026_TokenTimeBomb,
    author={Yang, Jingcheng and Wang, Enze and Chen, Jianjun and Wang, Qi and Zhang, Yuheng and Duan, Haixin and Xie, Wei and Wang, Baosheng},
  title     = {Token Time Bomb: Evaluating {JWT} Implementations for Vulnerability Discovery},
  booktitle = {Proceedings of the Network and Distributed System Security Symposium (NDSS)},
  year      = {2026},
  address = {USA},
  doi = {10.14722/ndss.2026.24069}
}

@misc{RFC6749,
  series       = {Request for Comments},
  number       = 6749,
  howpublished = {RFC 6749},
  publisher    = {RFC Editor},
  doi          = {10.17487/RFC6749},
  url          = {https://www.rfc-editor.org/info/rfc6749},
  author       = {Dick Hardt},
  title        = {{The OAuth 2.0 Authorization Framework}},
  pagetotal    = 76,
  year         = 2012
}

@misc{OIDC-Core,
  author       = {N. Sakimura and J. Bradley and M. Jones and
                  B. {de Medeiros} and C. Mortimore},
  title        = {{OpenID} Connect Core~1.0 Incorporating Errata Set~2},
  howpublished = {{OpenID Foundation Specification}},
  year         = {2023},
  url          = {https://openid.net/specs/openid-connect-core-1_0.html}
}

@misc{RFC7519,
  series       = {Request for Comments},
  number       = 7519,
  howpublished = {RFC 7519},
  publisher    = {RFC Editor},
  doi          = {10.17487/rfc7519},
  url          = {https://www.rfc-editor.org/info/rfc7519},
  author       = {Michael B. Jones and John Bradley and Nat Sakimura},
  title        = {{JSON Web Token (JWT)}},
  pagetotal    = 30,
  year         = 2015
}

@misc{RFC8693,
  series       = {Request for Comments},
  number       = 8693,
  howpublished = {RFC 8693},
  publisher    = {RFC Editor},
  doi          = {10.17487/rfc8693},
  url          = {https://www.rfc-editor.org/info/rfc8693},
  author       = {Michael B. Jones and Anthony Nadalin and Brian Campbell and John Bradley and Chuck Mortimore},
  title        = {{OAuth 2.0 Token Exchange}},
  pagetotal    = 27,
  year         = 2020
}

@techreport{NIST80063C,
  author      = {Paul A. Grassi and Justin P. Richer and Sarah K. Squire and
                 James L. Fenton and Ellen M. Nadeau and Naomi B. Lefkovitz and
                 Jamie M. Danker and Yee-Yin Choong and Kristen K. Greene and
                 Mary F. Theofanos},
  title       = {{Digital Identity Guidelines: Federation and Assertions}},
  institution = {National Institute of Standards and Technology},
  type        = {NIST Special Publication},
  number      = {800-63C},
  year        = {2017},
  doi         = {10.6028/nist.sp.800-63c}
}

@techreport{SAML-Core,
  author      = {{OASIS Security Services Technical Committee}},
  title       = {Assertions and Protocols for the {OASIS} Security
                 Assertion Markup Language ({SAML}) V2.0},
  institution = {OASIS},
  type        = {OASIS Standard},
  number      = {saml-core-2.0-os},
  year        = {2005},
  month       = mar,
  url         = {https://docs.oasis-open.org/security/saml/v2.0/saml-core-2.0-os.pdf}
}

@techreport{SAML-Profiles,
  author      = {{OASIS Security Services Technical Committee}},
  title       = {Profiles for the {OASIS} Security Assertion Markup Language ({SAML}) V2.0},
  institution = {OASIS},
  type        = {OASIS Standard},
  number      = {saml-profiles-2.0-os},
  year        = {2005},
  month       = mar,
  url         = {https://docs.oasis-open.org/security/saml/v2.0/saml-profiles-2.0-os.pdf}
}

@misc{Keycloak,
  author       = {{Red Hat}},
  title        = {{Keycloak}: Open Source Identity and Access Management},
  year         = {2024},
  howpublished = {\url{https://github.com/keycloak/keycloak}},
  note         = {Accessed: 2026-04-08}
}

@misc{Casdoor,
  author       = {{Casbin}},
  title        = {{Casdoor}: An Open-Source Identity and Access Management Platform},
  year         = {2024},
  howpublished = {\url{https://github.com/casdoor/casdoor}},
  note         = {Accessed: 2026-04-08}
}

@misc{Dex,
  author       = {{Dex IdP}},
  title        = {{Dex}: {OpenID} Connect {(OIDC)} Identity and {OAuth}~2.0 Provider},
  year         = {2024},
  howpublished = {\url{https://github.com/dexidp/dex}},
  note         = {Accessed: 2026-04-08}
}

@misc{Authentik,
  author       = {Jens Langhammer},
  title        = {{authentik}: The Authentication Glue You Need},
  year         = {2024},
  howpublished = {\url{https://github.com/goauthentik/authentik}},
  note         = {Accessed: 2026-04-08}
}

@misc{Vault,
  author       = {{HashiCorp}},
  title        = {{Vault}: Identity-Based Secrets and Encryption Management System},
  year         = {2024},
  howpublished = {\url{https://github.com/hashicorp/vault}},
  note         = {Accessed: 2026-04-08}
}

@misc{Zitadel,
  author       = {{ZITADEL}},
  title        = {{ZITADEL}: Identity Management Platform},
  year         = {2024},
  howpublished = {\url{https://github.com/zitadel/zitadel}},
  note         = {Accessed: 2026-04-08}
}

@misc{porta,
  title        = {{PORTA}: Artifact},
  year         = {2026},
  howpublished = {\url{https://github.com/dlgroupuoft/PORTA}},
}

@misc{Logto,
  author       = {{Silverhand Inc.}},
  title        = {{Logto}: The Better Identity Infrastructure for Developers},
  year         = {2024},
  howpublished = {\url{https://github.com/logto-io/logto}},
  note         = {Accessed: 2026-04-08}
}

@misc{RFC7515,
  series       = {Request for Comments},
  number       = 7515,
  howpublished = {RFC 7515},
  publisher    = {RFC Editor},
  doi          = {10.17487/RFC7515},
  url          = {https://www.rfc-editor.org/info/rfc7515},
  author       = {Michael B. Jones and John Bradley and Nat Sakimura},
  title        = {{JSON Web Signature (JWS)}},
  pagetotal    = 59,
  year         = 2015,
}

@inproceedings{MaierEtAl2016_OAuthMBT,
author = {Yang, Ronghai and Li, Guanchen and Lau, Wing Cheong and Zhang, Kehuan and Hu, Pili},
title = {Model-based Security Testing: An Empirical Study on OAuth 2.0 Implementations},
year = {2016},
isbn = {9781450342339},
publisher = {Association for Computing Machinery},
address = {New York, NY, USA},
url = {https://doi.org/10.1145/2897845.2897874},
doi = {10.1145/2897845.2897874},
booktitle = {Proceedings of the 11th ACM on Asia Conference on Computer and Communications Security},
pages = {651–662},
numpages = {12},
location = {Xi'an, China},
series = {ASIA CCS '16}
}

@inproceedings{HussainEtAl2018_LTEInspector,
  author       = {Syed Rafiul Hussain and
                  Omar Chowdhury and
                  Shagufta Mehnaz and
                  Elisa Bertino},
  title        = {LTEInspector: {A} Systematic Approach for Adversarial Testing of 4G
                  {LTE}},
  booktitle    = {25th Annual Network and Distributed System Security Symposium, {NDSS}
                  2018, San Diego, California, USA, February 18-21, 2018},
  publisher    = {The Internet Society},
  year         = {2018},
  url          = {https://www.ndss-symposium.org/wp-content/uploads/2018/02/ndss2018\_02A-3\_Hussain\_paper.pdf},
  bibsource    = {dblp computer science bibliography, https://dblp.org}
}

@inproceedings{ChenEtAl2023_SherlockOnSpecs,
author = {Chen, Yi and Tang, Di and Yao, Yepeng and Zha, Mingming and Wang, XiaoFeng and Liu, Xiaozhong and Tang, Haixu and Liu, Baoxu},
title = {Sherlock on specs: building LTE conformance tests through automated reasoning},
year = {2023},
isbn = {978-1-939133-37-3},
publisher = {USENIX Association},
address = {USA},
booktitle = {Proceedings of the 32nd USENIX Conference on Security Symposium},
articleno = {198},
numpages = {17},
location = {Anaheim, CA, USA},
series = {SEC '23}
}

\appendix 
\section*{Open Science} 

All artifacts needed to evaluate this paper's contributions are released~\cite{porta}.

\paragraph{Dependencies reviewers should be aware of.} Running the full pipeline end-to-end requires an OpenAI API key for the profile-extraction and sequence-generation phases; the combined cost for the three runs reported in the paper was approximately \$69.5 for the vulnerability discovery experiment (\S\ref{sec:eval:discovery}) and \$53.53 for the recall experiment (\S\ref{sec:eval:redetection}). Reviewers who do not wish to obtain an API key can still reproduce the execution and verdict phases by replaying the saved \texttt{sequences.json} files against the live Docker containers, which requires no LLM access, we mention how to execute the \texttt{replay} mode of \toolname in \texttt{README.md}.

\section*{Ethical Considerations} 

We follow the ACM CCS ethics guidelines.
All experiments target open-source software running in isolated
Docker containers under our control.
No production systems or user data were involved.
We have followed responsible disclosure practices for all findings reported in this manuscript and the accompanying artifact, completing the disclosure process with the relevant parties.

\section*{Generative AI Usage}

We disclose generative AI usage in this work in three categories.

\paragraph{LLM components within \toolname} The \toolname pipeline (\S\ref{sec:design}) uses an LLM (\texttt{gpt-5\_2-2025-12-1}, default temperature, top-$p=1.0$) for three internal phases: platform profile generation (\S\ref{sec:des:profile}), attack sequence generation (\S\ref{sec:des:generation}), and oracle assertion generation. Every LLM output is validated by deterministic execution against a live broker instance: profiles are subjected to majority-vote merging across three independent runs and validated against the platform before use; sequences and oracle assertions are evaluated by the deterministic verdict pipeline (\S\ref{sec:des:verdict}) over captured HTTP traces. No reported finding depends on the LLM's judgment alone.

\paragraph{AI-assisted source code lookup during triage.} During the manual triage of the 252 candidate findings (\S\ref{sec:eval:discovery}), we used Claude agents as a semantic search tool over each platform's source code. Given a captured HTTP request/response trace from a \toolname run, the agent located the source file and function relevant to the observed behavior. All root-cause analysis, vulnerability classification (specification violation versus under-specified flaw), assignment to invariant and code-level pattern, and the resulting 33-finding catalog were performed manually by the authors.

\paragraph{Writing assistance.} We used generative AI tools for grammar and style polishing of paper text. All technical content, claims, figures, tables, and arguments, that were improved or checked by AI tools, has been reviewed and verified by the authors before submission.

\section*{CVE numbers}
  \textbf{CVE-2026-15611}, \textbf{CVE-2026-15612}, CVE-2026-15615, \textbf{CVE-2026-15616}, \textbf{CVE-2026-15617}
    , CVE-2026-41577
    , CVE-2026-55669
    , CVE-2026-56664
     , CVE-2026-56665
     , CVE-2026-56666
     , \textbf{CVE-2026-9090}
     , CVE-2026-9091
     , \textbf{CVE-2026-9092}
     , \textbf{CVE-2026-9093}
     , \textbf{CVE-2026-9094}
     , CVE-2026-9095
    ,  CVE-2026-9096
     , \textbf{CVE-2026-9097}
    ,  \textbf{CVE-2026-9098} The bold entries indicate CVSS $\geq$ 9.0 (Critical). 

\end{document}